# Recalibration of the Sunspot Number: Status Report

F. Clette[a], L. Lefèvre[a], T. Chatzistergos[b], H. Hayakawa[c], V.M. Carrasco[d], R. Arlt[e], E.W. Cliver[f], T. Dudok de Wit[g], T. Friedli[h], N. Karachik[j], G. Kopp[j], M. Lockwood[k], S. Mathieu[l], A. Muñoz-Jaramillo[m], M. Owens[k], D. Pesnell[n], A. Pevtsov[f], L. Svalgaard[o], I.G. Usoskin[p], L. van Driel-Gesztelyi[q], J.M. Vaquero[d]

[a] World Data Center SILSO, Royal Observatory of Belgium, 3 avenue Circulaire, 1180 Brussels, Belgium

[b] Max Planck Institute for Solar System Research, Justus-von-Liebig-weg 3, 37077 Göttingen, Germany

[c] (1) Institute for Space-Earth Environmental Research, Nagoya University, Nagoya, 4648601, Japan; (2) Institute for Advanced Research, Nagoya University, Nagoya, 4648601, Japan; (3) Space Physics and Operations Division, RAL Space, Science and Technology Facilities Council, Rutherford Appleton Laboratory, Harwell Oxford, Didcot, Oxfordshire, OX11 0QX, UK; (4) Nishina Centre, Riken, Wako, 3510198, Japan

[d] Departamento de Física, Universidad de Extremadura, E-06006 Badajoz, Spain

[e] Leibniz-Institut f. Astrophysik Potsdam, An der Sternwarte 16, 14482 Potsdam, Germany

[f] National Solar Observatory, 3665 Discovery Drive, 3rd Floor, Boulder, CO 80303 USA

[g] (1) LPC2E, CNRS/CNES/University of Orléans, 45067 Orléans, France; (2) International Space Science Intititute, Hallerstrasse 6, 3012 Bern, Switzerland

[h] Rudolf Wolf Society, Ahornweg 29, 3123 Belp, Switzerland

[i] Ulugh Beg Astronomical Institute, Uzbekistan Academy of Sciences, 33 Astronomicheskaya str., 100052 Tashkent, Uzbekistan

[j] Laboratory for Atmospheric and Space Physics, University of Colorado, 3665 Discovery Dr., Boulder, CO 80303, USA

[k] Department of Meteorology, University of Reading, Earley Gate, PO Box 243, Reading RG6 6BB, UK

[l] ISBA/LIDAM, UCLouvain, Pl. de l'Université 1, 1348 Ottignies-Louvain-la-Neuve, Belgium

[m] Division of Solar System Science, Southwest Research Institute, Boulder, CO, USA

[n] Code 671, NASA Goddard Space Flight Center, Greenbelt, MD 20771, USA

[o] Hansen Experimental Physics Lab, Stanford University, Stanford, CA 94305, USA

[p] Space Physics and Astronomy Research Unit and Sodankylä Geophysical Observatory, Univeristy of Oulu, 90014 Oulu, Finland

[q] (1) University College London, Mullard Space Science Laboratory, Holmbury St. Mary, Dorking, Surrey, RH5 6NT, UK; (2) Observatoire de Paris, Université PSL, CNRS, Sorbonne Université, Univ. Paris Diderot, Sorbonne Paris Cité, 5 place Jules Janssen, F-92195 Meudon, France; (3) Konkoly Observatory, Research Centre for Astronomy and Earth Sciences, Konkoly Thege út 15-17, H-1121, Budapest, Hungary

**Abstract:** We report progress on the on-going recalibration of the Wolf sunspot number ($S_N$) and Group sunspot number ($G_N$) following the release of version 2.0 of $S_N$ in 2015. This report constitutes both an update of the efforts reported in the 2016 Topical Issue of Solar Physics and a summary of work by the International Space Science Institute (ISSI) International Team formed in 2017 to develop optimal $S_N$ and $G_N$ re-construction methods while continuing to expand the historical sunspot number database. Significant progress has been made on the database side while more work is needed to bring the various proposed $S_N$ and (primarily) $G_N$ reconstruction methods closer to maturity, after which the new reconstructions (or combinations thereof) can be compared with (a) "benchmark" expectations for any normalization scheme (e.g., a general increase in observer normalization factors going back in time), and (b) independent proxy data series such as F10.7 and the daily range of variations of Earth's undisturbed magnetic field. New versions of the underlying databases for $S_N$ and $G_N$ will shortly become available for years through 2022 and we anticipate the release of next versions of these two time series in 2024.

**1. Introduction**

The sunspot number ($S_N$; Clette et al., 2014, 2015; Clette and Lefèvre, 2016) is a time series (1700-present) that traces the 11-yr cyclic and secular variation of solar activity and thus space weather, making $S_N$ a critical parameter for our increasingly technological-based society (e.g., Baker et al., 2008; Cannon et al., 2013; Rozanov et al., 2019; Hapgood et al., 2021). While modern day observations may provide better parameters for characterizing the space weather effects, $S_N$ is the oldest direct observation of solar activity, and thus, is an indispensable bridge linking past and present solar behavior. As such, the sunspot number is the primary input for reconstructions of total solar irradiance (TSI) for years before 1940 (Wang et al., 2005; Krivova, Balmaceda, and Solanki, 2007; Kopp, 2016; Kopp et al., 2016; Wu et al., 2018b; Coddington et al., 2019; Wang and Lean, 2021).

The original formula of Rudolf Wolf (1816-1893; Friedli, 2016) for the daily sunspot number of a single observer is given by

$$S_N = k \left[ (10 \times G) + S \right] \qquad (1)$$

(Wolf, 1851, 1856), where G is the number of sunspot groups on the solar disk on a given day, S denotes the total number of individual spots within those groups, and k is a normalization factor that brings different observers to a common scale (k is the time-averaged ratio of daily sunspot number $S_N$ of the primary reference observer to that of a secondary observer). Because $S \approx 10\ G$ on average (Waldmeier, 1968; Clette et al., 2014), the two parameters have about equal weight in $S_N$.

G and S counts can vary between observers because of differences in telescopes, visual acuity, and environmental conditions. These factors are susceptible to both gradual (e.g., visual acuity) and abrupt (equipment-related) variations. Moreover, as we will see below, the working definitions of both G and S have changed over time. Getting the time-varying scaling relationships (k-coefficients) between multiple sets of overlapping observers correct over a time span of centuries – with frequent data gaps due to weather for individual observers and occasional periods of no overlap between any observers – is the daunting task of any reconstruction method. Finally, to add unavoidable noise to the process, there are differences between the daily G and S observations at different locations that are independent of instrumentation/environment/observer acuity and practice, reflecting only separation in universal time, either because of solar rotation (spots appearing at the Sun's east limb and/or disappearing at the west limb) as well as intrinsic solar changes, i.e., the evolution of spots and groups.

In order to illustrate the kind of differences that may appear between two sunspot observations, Figure 1 shows drawings made on the same day at two different stations. Figure 1 (a) is a drawing from the Specola Solare Ticinese station in Locarno, Switzerland, showing the delineation of groups and counting of spots on 14 March 2000 near the maximum of solar cycle 23 (1996-2008). Figure 1 (b) shows a sunspot drawing taken on the same date, about 7 hours later, at the US National Solar Observatory at Sacramento Peak in Sunspot, New Mexico (Carrasco et al., 2021a). While the drawings exhibit significant similarities, there are differences in the total number of sunspots and groups.

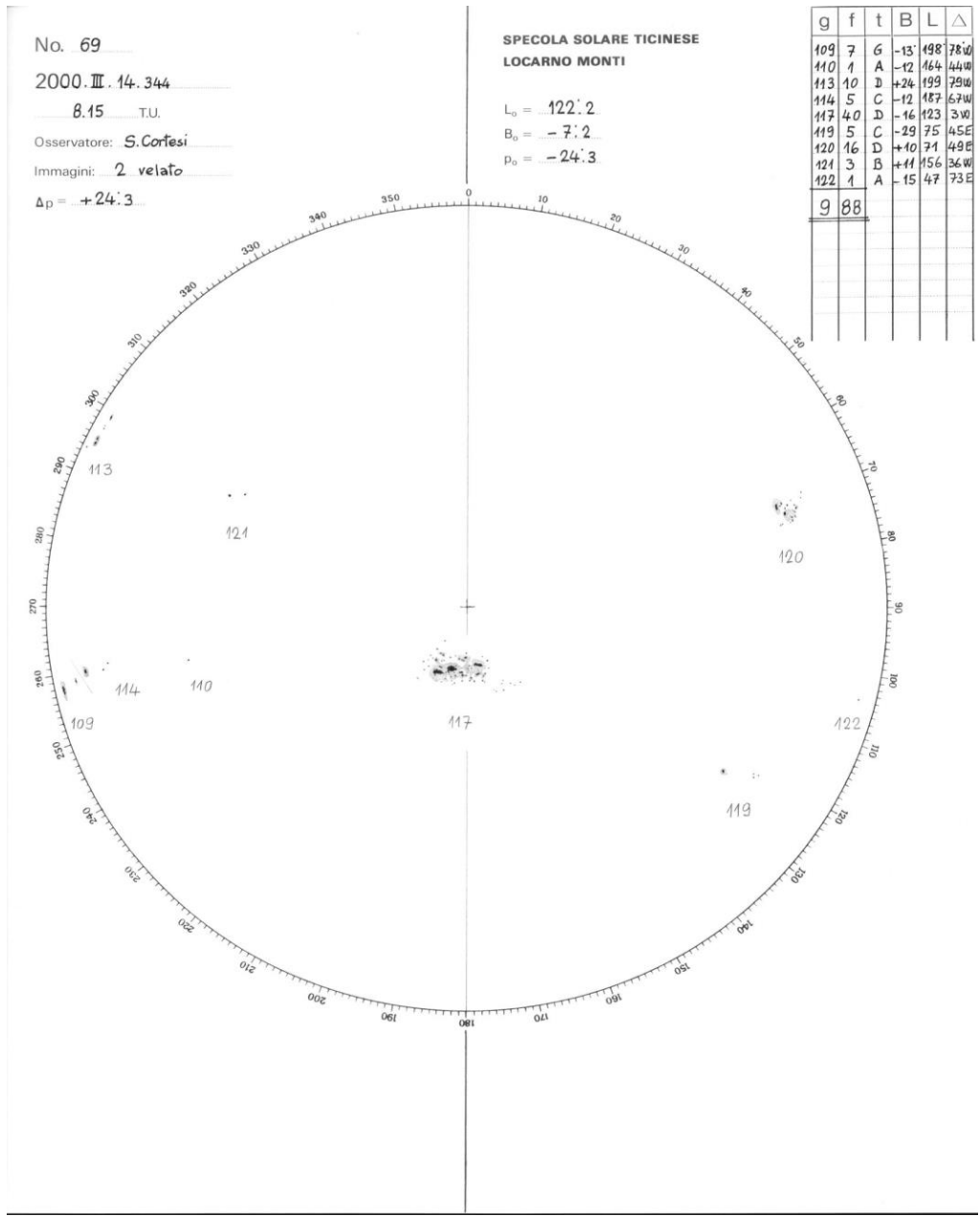

Figure 1. (a) Sunspot drawing by Sergio Cortesi from Specola Solare Ticinese on 14 March 2000. Nine groups and 88 individual spots (in the table at the top right corner marked as "flecken" which is "spots" in German) were observed to yield SN (Locarno) =178. (https://www.specola.ch/e/drawings.html)

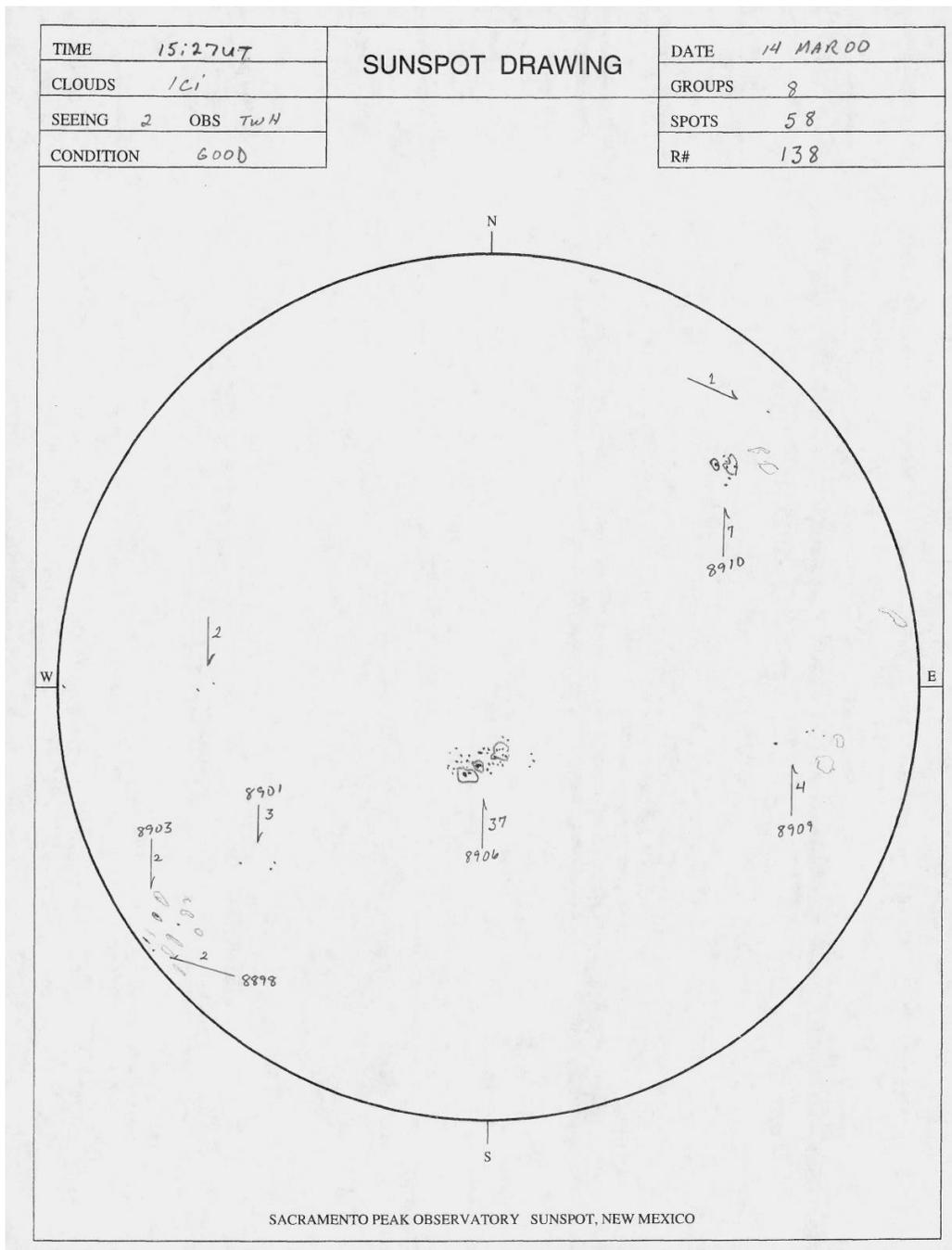

Figure 1. (b) Sunspot drawing by Tim Henry from the US National Solar Observatory at Sacramento Peak on 14 March 2000, about 7 hours after the Locarno observation in Figure 1(a). Eight groups and 58 sunspots were observed to yield SN (SacPeak) = 138. Fainter contours outline photospheric faculae, visible near the solar limb. (https://ngdc.noaa.gov/stp/space-weather/solar-data/solar-imagery/photosphere/sunspot-drawings/sac-peak/)

Although the difference in the number of groups between two drawings is minimal (8 vs. 9), there are several other differences: e.g., groups 113 and 122 shown in Locarno observations are missing from Sacramento Peak drawings, and an unnumbered group to the northwest of AR 8910 that consists of a single pore is not present in the Locarno drawing. Just this single spot thus produces a difference of 11 in $S_N$. Groups 113 and 122 and the unnumbered group northwest of AR 8910 were all located close to solar limbs where visibility effects could be strongest. Such discrepancies can thus result from the difference in the time at which the observations were made, as a result of intrinsic changes on the Sun, viz., sunspot appearance or disappearance, growth or decay, as well as solar rotation. A study of the random noise in $S_N$ (Dudok de Wit et al, 2016) shows that the random evolution of solar active regions dominates over observing errors in the discrepancies between observations for the same date. Overall, in this case, $S_N$ (Locarno; 14 Mar 2000) = 178 while $S_N$ (SacPeak; 14 Mar 2000) = 138.

Wolf modified the $S_N$ time series that he had initiated several times before his death in 1893, but only for the years before 1848 when he began observing, i.e., for recovered data from early observers (see Section 2.1 in Clette et al., 2014). Wolfer, Wolf's successor as observatory director at Zürich, made a final revision to $S_N$ in 1902 (for the 1802-1830 interval) based on the addition of new data from Kremsmünster. Thereafter, the $S_N$ series, as maintained by the Zürich Observatory until 1980 and subsequently by the Royal Observatory of Belgium (ROB), in Brussels, was left unchanged but continuously extended by appending new data to keep it up to date until the present.

$S_N$ remained the only long-term time series directly retracing solar activity until 1998, when Hoyt and Schatten (1998a,b) developed a group sunspot number ($G_N$)

$$G_N = k'G \qquad (2)$$

where k' is the station normalization factor. This simpler index does not include the count of individual spots, which characterizes the varying size of the sunspot groups but is often missing in early observations.

A key advantage of the $G_N$ series was that Hoyt and Schatten (1998a,b) were able to reconstruct it back to the first telescopic observations of sunspots by Galileo and others ca. 1610 vs. the initial year of 1700 for Wolf's $S_N$ series. The new $G_N$ thus encompassed the period of extremely weak sunspot activity from 1645-1715 known as the Maunder Minimum (Eddy, 1976; Spörer, 1887, 1889; Maunder, 1894, 1922). Until recently, another crucial advantage of $G_N$ was the availability of all the raw source data in the form of an open digital database, while the $S_N$ source data, forming a much larger collection (more than ~800,000 observations), were available in digital format only since 1981 (Brussels period). Older $S_N$ data existed only in their original paper form or were even long considered as partly lost (see Section 2.1.1 below). The unavailability of those core data has so far prevented a full end-to-end reconstruction of the original $S_N$ series from its base elements.

The $S_N$ parameter defined by Wolf in Equation 1 requires more detailed observations than $G_N$ because of the inclusion of the S (spot) component. However, early observations are often ambiguous, as many observers did not make a clear distinction between spots and groups, using the same name "sunspot" or "kernel" for both. Moreover, as can be seen in Figure 1, large spots are enveloped in a penumbra that may contain one or several dark umbrae. Depending on the observer and epoch, each penumbra will either be counted as 1 in the S count, regardless of the number of embedded umbrae, or all umbrae will be counted separately, leading to a higher S value. Both the S and G counts are affected by the acuity threshold problem faced by each observer of distinguishing the smallest spots, in particular, for small groups consisting of a solitary spot (for which G = 1, S = 1, and $S_N$ = 11) and versus the absence of spots (G, S, and $S_N$ = 0). Finally, the G variable presents its own difficulty: the separation of a cluster of magnetic dipoles into one or more distinct groups, depending on the cluster morphology and evolution. This splitting of groups depends on personal practices and on the scientific knowledge at different epochs, and becomes difficult mainly for a heavily spotted Sun, near the solar cycle maximum.

1.1 Motivation for the ISSI International Team on recalibration of the sunspot number

The series $G_N$ and $S_N$[1] can be considered to be equivalent if one assumes that S is proportional to G, on average, with a constant of proportionality that does not change over time. In this case, differences between the form of the $G_N$ and $S_N$ time series would indicate calibration drifts in one or both time series. However, the possibility of a long-term change in solar behavior would mean that the ratio S/G could have varied and this would be a separate cause of deviations of $G_N$ from $S_N$. A modulation of the S/G ratio by the solar cycle has actually been found by Clette et al. (2016b) and Svalgaard, Cagnotti and Cortesi (2017), indicating a non-linear relation between the two indices that seems to be stable in time. Therefore, where they overlap, $S_N$ and $G_N$ can be considered as close equivalents but with different detailed properties and calibrations.

While the original Zürich $S_N$ (termed $S_N$ (1) herein) and Hoyt and Schatten (1998a,b; HoSc98) $G_N$ series agreed reasonably well over their 1874-1976 normalization interval, HoSc98 was significantly lower for maxima before ~1880 (see Figure 1(a) in Clette et al., 2015, and Figure 2 (below)). This situation was puzzling, if not unacceptable. What was the cause of the divergence? Could the two series be reconciled? Such questions have led to a decade-long effort by the solar community to construct more homogeneous and trustworthy $S_N$ and $G_N$ time series beginning with a sequence of four Sunspot Number Workshops from 2011-2014 (Cliver, Clette, and Svalgaard, 2013a; Cliver et al., 2015). This effort produced major recalibrations of both $S_N$ (1) by Clette and Lefèvre (2016) to yield $S_N$ (2) and HoSc98 by Svalgaard and Schatten (2016) to yield SvSc16.

These recalibrations removed the marked divergence of $S_N$ (1) and HoSc98 before ~1880 (see Figure 1(b) in Clette et al., 2015) and reduced differences during the 18[th] century (Clette et al., 2015). However, questions were raised about the validity of the methods used (Usoskin et al., 2016a; Lockwood et al., 2016) and new ideas for recalibration emerged. The revisions of the Wolf $S_N$ series and the HoSc98 $G_N$ series were accompanied by an independent revision/extension of $S_N$ (Lockwood, Owens, and Barnard, 2014a,b; Lockwood et al., 2016; LEA14) and novel reconstructions of $G_N$ by Usoskin et al. (2016a; UEA16) and Chatzistergos et al. (2017; CEA17). Modifications to UEA16 was published by Willamo et al. (2017; WEA17) and WEA17 was subsequently modified by basing it on the new Vaquero et al. (2016; V16 herein; 16 for the year of release) database[2] rather than the original Hoyt and Schatten (1998) data base (Usoskin et al., 2021; UEA21). The resultant situation, documented in a Topical Issue of Solar Physics (Clette et al., 2016a), was similar to that which motivated the Sunspot Number Workshops, but writ larger (Cliver, 2017;

---

[1] Throughout this update, we will use $S_N$ and $G_N$ generically for composite time series of sunspot numbers and group numbers irrespective of the calibration approach.

[2] Both the Vaquero et al. (2016; V16) data base for $G_N$ reconstruction and the original Hoyt and Schatten (1998a,b) observer data base can be found at https://www.sidc.be/silso/groupnumberv3.

Cliver and Herbst, 2018): several independently constructed series that diverged before ~1880 (Figure 2) – largely bounded by those of SvSc16 and HoSc98. This state of affairs prompted a successful International Space Science Institute (ISSI) Team proposal in 2017 entitled "Recalibration of the Sunspot Number Series" by Matt Owens and Frédéric Clette. Team meetings were held in Bern in January 2018 and August 2019. The ultimate goal, yet to be met, of the proposal was to provide consensus "best-method" reconstructions of $S_N$ and $G_N$ including quantitative time-dependent uncertainties, for use by the scientific community. Here we report the results of this effort and provide an update of the 2016 Topical Issue.

In Section 2, for $S_N$ and $G_N$, in turn, we present highlights of the data recovery effort and discuss work on sunspot number reconstruction methodologies. In Section 3, we discuss two "benchmarks" for sunspot number recalibration (quasi-constancy of spot-to-group ratios over time and an increase, as one goes back in time, of k- and k'-values). In Section 4, we review associated work on independent long-term measures of solar activity. Such proxies as quiet-Sun solar radio emission, solar-induced geomagnetic variability, and cosmogenic radionuclide concentrations, can be used as correlates/checks on new versions of $S_N$ and $G_N$. We stress, however, that, thus far, the sunspot and group number back to 1610 remain fully independent of these proxies. In section 5, we discuss efforts to connect primary observers Schwabe and Staudacher across the data-sparse interval from 1800-1825 and the challenges of the first ~140 years of $S_N$ and $G_N$ time series (1610-1748). Sections 6 and 7 contain a summary of ISSI Team results and a perspective/prospectus for the on-going recalibration effort, respectively.

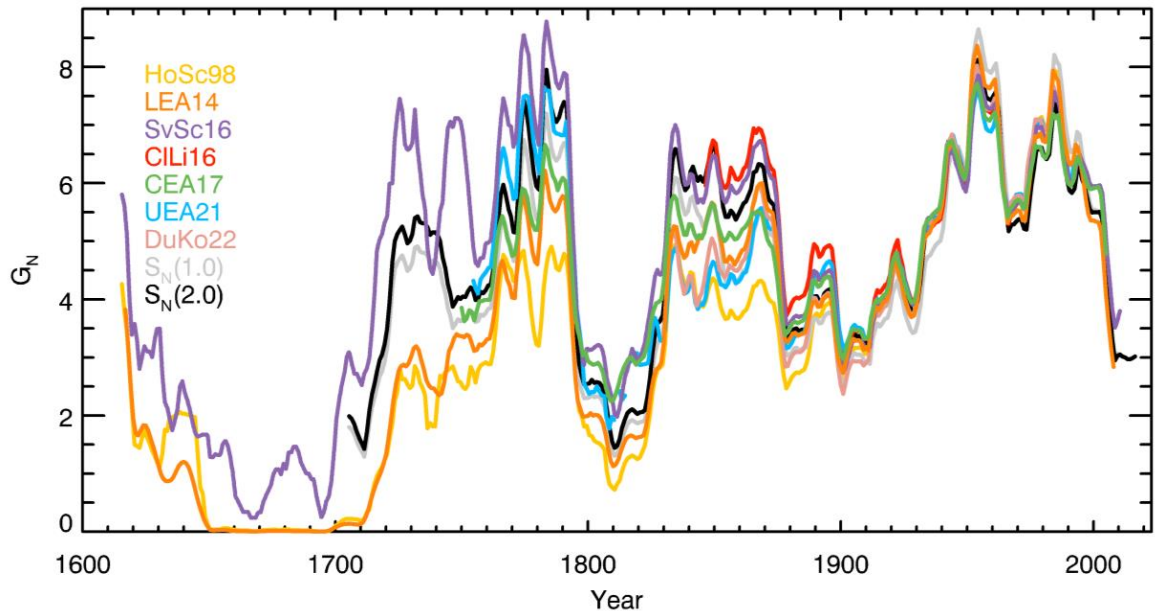

Figure 2. Eleven-year running means of eight sunspot series: The original Wolf $S_N$ ($S_N(1.0)$) series, the Hoyt and Schatten (1998a,b) $G_N$ (HoSc98) series, the Lockwood et al. (2014a,b) $S_N$ (LEA14) series, the Clette and Lefèvre (2016) corrected $S_N$ ($S_N(2.0)$) series, the Svalgaard and Schatten (2016) reconstructed $G_N$ (SvSc16) series, the Cliver and Ling (2016) $G_N$ (ClLi16) series, the Chatzistergos et al. (2017) $G_N$ (CEA17) series, and the Usoskin et al. (2021a) $G_N$ (UEA21) series with predecessors given in Usoskin et al. (2016a; UEA16) and Willamo et al. (2017; WEA17). All records are scaled to the mean $S_N(2.0)$ series over 1920-1974.

**2. Solar Physics Topical Issue update / ISSI Team report**

## 2.1 Wolf sunspot number ($S_N$)

*2.1.1 Data Recovery: the Sunspot Number Database*

Data recovery is a never-ending task, and future revisions of sunspot series will occur intermittently as part of a continuous upgrading process. The last few years have witnessed significant advances in recovery and digitization of data underlying the $S_N$ time series, particularly in the context of the ISSI Team collaboration over 2019-2021. The focus in this section is on sunspot observations that were either made at Zürich and later at Locarno and Brussels for the creation of $S_N$, or collected and archived from external observatories by the Directors of the Zürich Observatory, and since 1981 by the World Data Center for Sunspot Index and Long-term Solar Observations (WDC-SILSO) in Brussels, for this purpose.

The Zürich sunspot number produced by Wolf and his successors from 1700-1980 is based on three types of data: (1) the raw counts from the Zürich observers (the director and the assistants) from 1848-1980, (2) corresponding counts sent to the Observatory of Zürich by external auxiliary observers, and (3) historical observations for years before 1849 collected mainly by Wolf but also by A. Wolfer, his successor. Much of this material was published over the years in the bulletins of the Zürich Observatory, the Mittheilungen der Eidgenössischen Sternwarte Zürich (hereafter, the Mittheilungen). This is a fundamental resource for any future re-computation of $S_N$.

Until recently, this large collection of printed data was completely inaccessible in digital form. During 2018-2019, a full encoding of the Mittheilungen data tables, from the printed originals, was undertaken at WDC-SILSO in Brussels, constituting the initial segment of the sunspot number database. This includes all data published between 1849 and 1944 (black curve in Figure 3) when Max Waldmeier, the last Director of the Zürich Observatory decided to cease publishing raw data. This database now contains 205,000 individual daily sunspot counts (Clette et al., 2021): it currently includes data from the Zürich and auxiliary observers between 1849 and 1944, and all long data series in the early historical part, from 1749 to 1849, that were found and collected by R.Wolf in his epoch. (NB: Isolated observations, randomly scattered over time, are less exploitable and will be added later.) In addition to daily counts of spots and groups, the database includes metadata - in particular, changes of observers or instruments.

Although the published data were globally comprehensive, some parts are missing. The Mittheilungen provide the full set of observations from the Zürich team itself only from 1871 onwards, when Wolf started to publish separate tables for himself and his assistants as well as data received from external observers (Friedli, 2016). Before 1871, Wolf published only a single yearly table, with his all his own observations, and data from external observers or assistants were only included to fill the remaining random missing days. Later on, after World War I (WW I), Alfred Wolfer added many external observers, creating a truly international network. However, given the strong increase in the amount of collected data, for financial reasons, he greatly reduced the published raw data, limiting them to those from Zürich observers and ~10 external observers. When William O. Brunner succeeded Wolfer as director of the Zürich Observatory in 1926, he stopped publishing data from external observers. Only observations from the Zürich Observatory itself and from Karl Rapp (Locarno, Switzerland) were published after that year. Then in 1945, when Max Waldmeier became the new Director, the publication of source data ceased completely, and only a list of contributing stations was provided for each year.

All those unpublished data, collected during the Zürich era, were stored in paper archives at the Zürich Observatory, which were supposed to include the whole collection, though in a less directly accessible manner. Unfortunately, following the closing of the Observatory of Zürich in 1980, when the curatorship of the sunspot number was transferred to ROB, those unpublished archives went missing. Figure 3 (brown and blue curves) shows the resulting major ~60-year shortfall (1919-1979) in the preserved raw data. The absence of a large portion of the 1919-1979 sunspot data constituted a critical missing link between the early Zürich epoch and the modern international sunspot number produced in Brussels since 1981, for which all data are preserved in a digital database[3]. In particular, this period spans one of the main scale discontinuities identified in the Zürich series, the "Waldmeier jump" in 1947 (Clette et al., 2014; Lockwood, Owens, and Barnard, 2014a; Clette and Lefèvre, 2016; Svalgaard, Cagnotti, and Cortesi, 2017).

In a key development, the amount of missing data from 1919-1980 was greatly reduced in late 2018 and early 2019 when a serendipitous find by the staff of the Specola Solare Ticinese Observatory in Locarno, followed by subsequent searches in Zürich and Brussels, resulted in the recovery of the entire Waldmeier data archive (1945-1979) (Clette et al., 2021). The recovered Waldmeier data archive includes the last 35 years of the Zürich era, up to the transition with WDC-SILSO in Brussels, in the form of yearly handwritten tables. They thus bring the essential missing link between the entire past Zürich series and the current SILSO $S_N$ production and its complete input-data archive. The recovered paper originals have now been converted to digital images by the ETH Zürich University Archives (ETH catalogue entry: Hs1304.8) and are accessible on-line on the Swiss e-manuscripta portal (https://www.e-manuscripta.ch).

This recovery already delivered key explanations regarding the main discontinuity formerly found in 1947 in the $S_N$ series (Clette et al., 2021; see below), but the time-consuming digitization of the huge set of observations (more than 300,000 daily numbers) to expand the $S_N$ database will require more resources and time. Simultaneously, efforts are continuing to recover the last missing data from the auxiliary stations between 1919 and 1944. Figure 3 shows that the recently digitized data from the unpublished source data of the Mittheilungen constitutes only about half of the observations taken and collected from external sources by the Zürich Observatory from the end of World War I until 1980.

---

[3] 1980 was a disturbed closing year at the Zürich Observatory. The standard yearly tables of source data were not produced by the Zürich staff, and the resulting sunspot numbers were published as crude typewritten pages instead of the normal printed edition of the Mittheilungen. However, the original reports sent to Zürich by all auxiliary stations were recovered and were transferred in 2007 to the World Data Center SILSO in Brussels. Those original reports were exploited for the production of the revised series $S_N$ V2.0 in 2015, as an important link to join the Zurich series and the international sunspot number across the 1980-1981 transfer from Zürich to Brussels. As a consequence of this disorganization in 1980, the collection of standard Zürich source tables thus ends in 1979. All internal data from Zürich observers for 1980 as well as the documentation about the monthly $S_N$ processing were provided to the new team in Brussels by the assistants of the Zürich observatory at the occasion of personal visits in Zürich and Brussels.

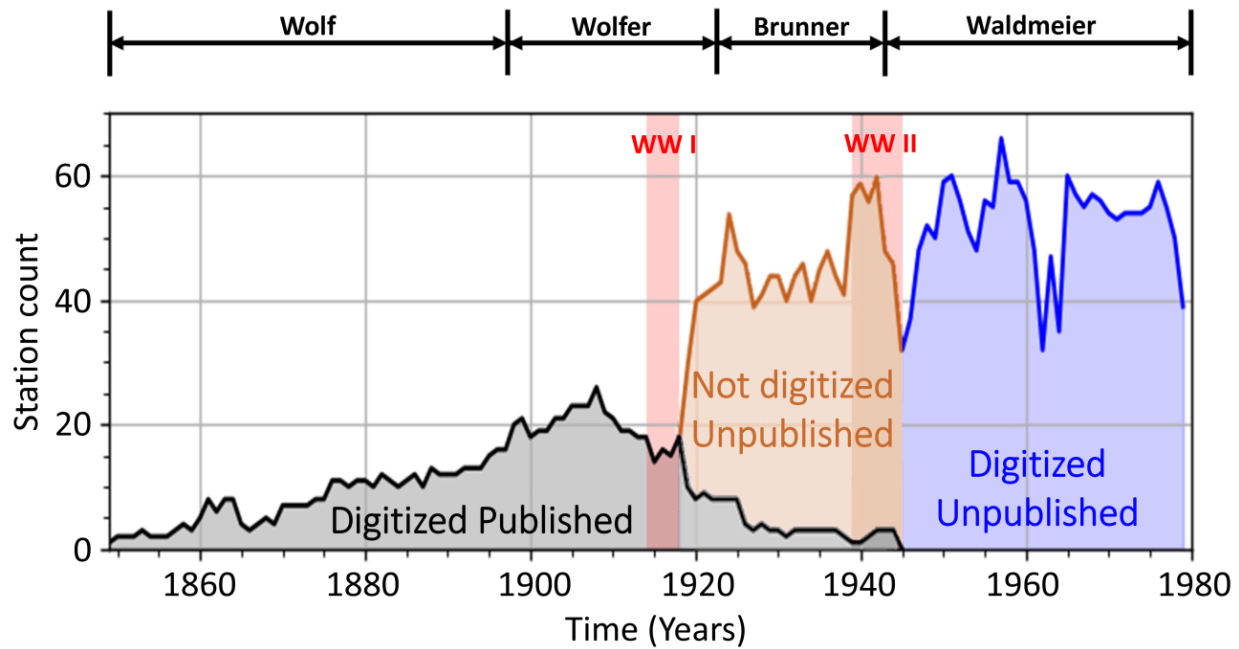

Figure 3. Evolution of the number of contributing stations collected by Zürich for each year. The raw source data published in the Mittheilungen of the Zürich Observatory (gray curve) have now been digitized into a new database.  After 1918, the number of external stations grew significantly, but the Zürich Observatory did not to publish all its source data anymore, and even ceased completely after 1945. Those unpublished data, stored in handwritten archives but missing until recently, are shown by the brown and blue-shaded curves. In 2019, the archives for the period 1945-1979 were finally recovered (blue part), and the original sheets were scanned, but the values still need to be extracted to fill the database. The archived source data from 1919-1944 (brown part) are still missing, and are the target of further searches. The two vertical shaded bands mark the two World Wars, both of which left a clear imprint on the Zürich data set. Tenures of the Zürich Observatory Directors are indicated at the top of the panel. (Figure adapted from Clette et al., 2021.)

Finally, two important source documents also bring essential information regarding the early part of Wolf's period, between 1849 and 1877: Wolf's handwritten source books and the collection of input-data tables maintained first by Wolf and then by Wolfer until 1908 (Friedli, 2020). Both documents are preserved at the ETH Zürich University Archives, and have now been entirely scanned and made accessible online. Wolf's source books provide yearly tables and contains invaluable information about the calculation of each daily sunspot number, which was never published in the Mittheilungen: e.g., the distinction between Wolf and Schwabe's daily numbers, where they overlapped between 1849 and 1868, and yearly k coefficients for each external observer. The source tables are more comprehensive and list all raw counts from all observers, back to 1610, including data that were never used for production of the sunspot number.  So far, part of the source book has been encoded (1849-1877) by the Rudolph Wolf Society, and is accessible on-line (Friedli, 2016; www.wolfinstitute.ch). Those original data complement the published tables in the Mittheilungen, and will prove essential to better understand the multiple methodological changes introduced by Wolf during his long career – in particular, the scale transfers between Schwabe and Wolf over 1849-1868 and between Wolf and Wolfer over 1877-1893 (Friedli 2020, see below; Bhattacharya et al., 2022, submitted).

*2.1.2 Reconstruction Methodology for $S_N$*

*2.1.2.1 The new $S_N(2.0)$ series*

Two primary $S_N$ time series have been constructed thus far: the original series $S_N$ (version 1.0; 1700-2014) constructed by Wolf (1851, 1856) and its first revised version $S_N(2.0;$ 1700-present) (Clette and Lefèvre, 2016). $S_N(1.0)$ and $S_N(2.0)$ give daily values starting in 1818, monthly values starting in 1749 and yearly averages starting in 1700. Both series are available at http://www.sidc.be/silso/. In addition, Lockwood, Owens, and Barnard (2014a,b) constructed a $S_N$ series (LEA14; 1610-2012) by appending a scaled-up Hoyt and Schatten (1998a,b) $G_N$ series for years prior to 1749 to $S_N(1.0)$ with corrections applied ca. 1850 and 1950, using geomagnetic indices as an external reference. Friedli (2016, 2020; $F(S_N(1.0)_{COR})$) reconstructed the 1877-1893 segment of $S_N(1.0)$ for which Wolf and Wolfer overlapped. Specifics of these four series are summarized in Table 1.

Table 1. $S_N$ time series

| Ref. | Abbrev. | Years | Cadence | Source | Calibration Stitching |
|---|---|---|---|---|---|
| 1 | $S_N(1.0)$* | 1818-2015 | daily | Zürich/SILSO | pilot observer (single station), linear scaling, daisy-chain and external index before 1848 |
| 2 | $S_N(2.0)$ | 1818-2020 | daily | Zürich/SILSO | pilot observer (single and multi-station), linear scaling |
| 3 | LEA14 | 1818-2014 | daily | Lockwood et al. (2014b; as supporting information) | linear scaling, external indices provisional** |
| 4 | $F(S_N(1.0)_{COR})$ | 1877-1893 | monthly | Friedli (2020; Table 2) | linear scaling of Wolf to Wolfer, time-limited correction to $S_N(1.0)$ |

References: 1 = (Wolf, 1851, 1856; see Section 2.1 in Clette et al. (2014) for Wolf's subsequent development of $S_N(1.0)$), Wolfer (1902), Waldmeier (1961), McKinnon (1987)); 2 = Clette and Lefèvre (2016); 3 = Lockwood, Owens, and Barnard (2014a,b); 4 = Friedli (2016, 2020)

* For the 1849-1980 period when the Wolf sunspot number series was produced at Zürich, it was variously referred to as $R_W$ (for Wolf) or $R_Z$ (for Zürich), where the R denotes a relative rather than absolute scale. After 1980, the Wolf series was commonly referred to as $R_i$ (for international). The $S_N$(version) notation used generically in this status report for the Wolf sunspot number series was introduced in Clette and Lefèvre(2016), when releasing the first re-calibrated series $S_N(2.0)$.

** Provisional in nature and not intended as a prescription or method for reconstructions of $S_N$

Version 2.0 of the $S_N$ time series (Clette and Lefèvre, 2016; $S_N(2.0)$) included three corrections applied to different time intervals. In reverse time order, these were: (1) the Locarno drift correction (1981-2015), (2) the Waldmeier jump (1947-1980), and (3) the "Schwabe-Wolf" correction (1849-1863). Those three corrections were based on the three largest anomalies diagnosed in the series.

In addition, a new conventional scale was adopted, taking Wolfer as reference observer instead of Wolf. This removed the 0.6 scale factor applied so far by heritage to all modern data after 1893, and it brought $S_N(2.0)$ in line with what is actually observed on the Sun with modern instruments. As this 0.6 downscaling factor uses partly undocumented early observations as a reference, and as it rests on assumptions and interpretations that may be questioned in the future (see next section), its application to data posterior to this less documented early period was inappropriate. In $S_N(2.0)$, the scale of all data since 1894 is now left unchanged, and instead, the inverse factor 1/0.6 (= 1.667) is applied to the early part of the series, before 1894, which is effectively the part for which the counts are incomplete and require a compensating factor.

Of the three above-mentioned corrections in $S_N(2.0)$, only the first one led to a full recalculation of the $S_N$, based on the raw input data, which are fully archived in digital form by SILSO since 1981 (Clette et al., 2016a, 2016b). The other two corrections consisted in the application of a correction factor to the original $S_N(1.0)$ Zürich series over the time interval affected by each diagnosed inhomogeneity (Clette and Lefèvre, 2016). Indeed, in the Zürich system before 1981, most of the daily sunspot numbers, as published in the Mittheilungen and in Waldmeier (1961), were simply the raw Wolf numbers (k coefficient fixed at unity) from the primary observer at the Zürich observatory, without any kind of statistical processing (Clette et al. 2014; Friedli 2016, 2020). In this scheme, the numbers from external auxiliary stations were only used on days when the Sun could not be observed in Zürich (on average, less than 20% of all days), and thus played a secondary role. This approach thus rested entirely on the personal life-long stability of individual reference observers, and on the assumed equivalence "by construction" of successive observers observing from the same place with mostly the same instrument. Any correction thus implies the use of alternate observers. Unfortunately, as described in the previous section, until recently, the Zürich source data were not available in digital form and part of the archives were even missing, which prevented any reconstruction from a wide base of source data from multiple alternate observers.

Reconstructing a sunspot database for the entire Zürich period before 1980 will thus be a key step for any future upgrade of the $S_N$ series. This focused the efforts over the last few years, as described in the previous section and also in Section 2.2.1. Although the database is still incomplete at this time and a full end-to-end reconstruction cannot be undertaken yet, the data gathered thus far already allowed to clarify two primary scale transitions in the $S_N$ series.

*2.1.2.2 The Wolf-Wolfer scale transfer*

The Wolfer to Wolf conversion factor of 0.6, embedded in the $S_N(1.0)$ time series, was introduced because Wolfer, Wolf's successor, counted more groups and spots than Wolf in simultaneous observations spanning 17 years (1877-1893). Those higher counts resulted from two simultaneous causes. First, Wolfer used the standard telescope of Zürich Observatory (Aperture: 83 mm; magnification 64x), while Wolf was only using a much smaller portable telescope during that part of his observing career (Aperture: 40mm, Magnification: 20x). This enabled Wolfer to see and count single small pores (small sunspots without penumbra and area from 0.5-4 millionths of a solar hemisphere (μsh); Tlatov et al., 2019) that were undetectable in Wolf's small telescope. In addition, as Clette et al. (2007) wrote: *"In 1882, A. Wolfer … introduced an important change in the counting method (Hossfield, 2002 [see also Wolf 1857; Wolfer 1895; Kopecký, Růžičková-Topolová, and Kuklin, 1980]). While Wolf had decided not to count the smallest sunspots visible only in good conditions and also not to take into account multiple umbrae in complex extended penumbrae, in order to better match his counts with the earlier historical observations, the new index included all small sunspots and multiple umbrae. [Figure 1 illustrates how small groups with low spot counts balance the effect of large groups with many spots in $S_N$.] By removing factors of personal subjectivity, this led to a much more robust definition of the $S_N$ that formed the baseline for all published counts after 1882. To complete this transition, A. Wolfer determined the scaling ratio between the new count and the Wolf SN series over the 16-year Wolf–Wolfer overlap period (1877–1892). This led to the constant Zürich reduction coefficient ($K_Z$ = 0.6) [that was used] to scale … [$S_N(1.0)$] to the pre-1882 Wolf sunspot counts."* (This change does not, however, affect the $G_N$ where multiple umbrae within the same penumbra do not alter the number of groups.) However, already in Wolfer's original study (Wolfer 1895), the yearly mean Wolf-Wolfer ratio shows a clear drift over the interval 1877-1883, followed by a stabilization (Figure 4, left panel), which sheds doubts on the accuracy of this mean 0.6 factor and indicates that it should be re-determined. Earlier studies confirmed the specificity of this transition (Svalgaard, 2013; Cliver and Ling, 2016; Cliver, 2017; Bhattacharya et al., 2021).

Recently, in the framework of the ISSI Team, Friedli (2020) revisited this important transition, by using unpublished source data tables compiled by Wolfer (1912) in the framework of a PhD thesis by his student Elsa Frenkel (1913; with Albert Einstein as supervisor). Those tables include data from multiple observers over the same time interval, allowing to retrace any drift in Wolf's or Wolfer's data (Figure 4, right panel). Friedli concludes that Wolfer was stable over this interval of overlap, while Wolf's series changed relative to all other observers, probably due to a slow degradation of his eyesight associated with ageing. In particular, he found that the mean factor should be lowered to 0.55, which means that the maxima of Cycle 12 (1884) and Cycle 13 (1893), which are framed by the 1877-1893 interval, should be ~10% higher than in the original Wolf series, as also suggested by Cliver and Ling (2016), implying a similar correction for $S_N$(2.0).

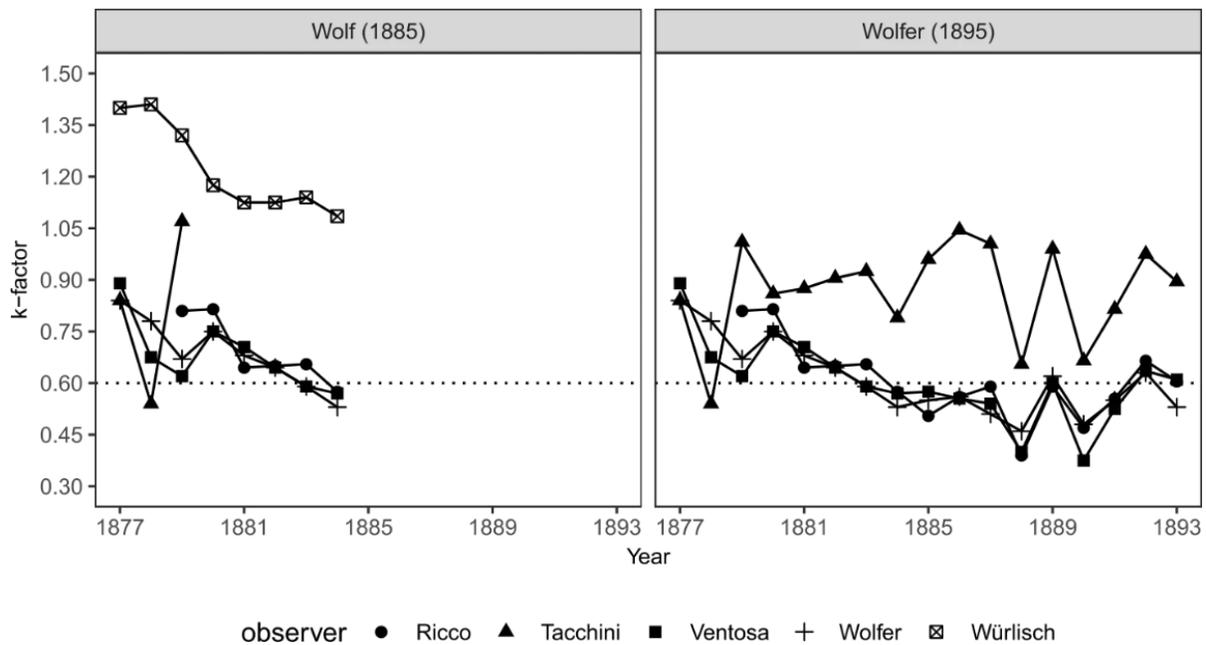

Figure 4. Left panel: yearly k-factors of Riccó, Tacchini, Ventosa, and of Wolfer relative to Wolf as given by Wolfer (1895), showing a clear drift before 1884. Right panel: yearly k-factors of Riccó and Ventosa as given by Frenkel (1913) using Wolfer as reference. The k factors are stable over the whole time interval, indicating that all observers are stable. By contrast, the yearly k-factor of Rudolf Wolf, with the small portable refractor relative to the numbers from Wolfer using the 83 mm standard Zürich refractor (squares) as given in the upper part of the panel, shows the same trend between 1877 and 1884, but also a modulation that follows the solar cycle (minima in 1879 and 1890, maxima in 1884 and 1894) (from Friedli, 2020).

Moreover, in the early part of the interval, between 1877 and 1883, the scale factor is lower than in the final part, after 1890, by up to a factor of 0.76. This would suggest that all $S_N$ values before 1877 were scaled too high by up to 25% relative to all $S_N$ values after 1894. However, the first years of this interval fall in the minimum between solar cycles 11 and 12. The uncertainty in the ratios between observers is thus very high, due to the low sunspot counts during that period. Moreover, Friedli's analysis shows that the comparison between observers with widely different personal k coefficients shows a significant modulation with the solar cycle, which could be the consequence of a non-linear relation between the counts of two observers with very different instruments. This is precisely the case for Wolf, who used only his small portable telescope, by contrast with all other observers who had much larger instruments, with aperture of 80mm or more. Therefore, Wolf's drift diagnosed here with a linear model may include, at least partly, a spurious solar cycle variation (Figure 4, right panel). In that case, the drift found in the rising part of cycle 12 (1877-1883) may actually reverse and vanish before 1877, when solar activity reached higher levels in cycle 11, as well as in the other cycles before it. Such possible effects were not analyzed yet, and Friedli (2020) rightly concludes: *"Before 1877, the scale transfer from the 40/700 mm Parisian refractor as used by Rudolf Wolf to the 83/1320 mm Fraunhofer refractor as used by Alfred Wolfer will need to be analyzed further."*

*2.1.2.3 The 1947 jump and the sunspot weighting effect*

By a comparison with data from the Madrid and Greenwich observatories, Svalgaard (2012, 2013) identified a sharp upward jump in the Zürich sunspot number, occurring around 1945-1947. A likely cause for this jump was quickly identified: the introduction in the observing practice of Zürich observers of a weighting of the sunspot counts according to the size of the spots (Svalgaard, 2013, Svalgaard, Cortesi and Cagnotti, 2017, Clette et al., 2014). The effects of this weighting can be seen in Figure 1 (a) for spot groups 114 (3 spots observed; 5 listed) and 121 (2 observed; 3 listed). In order to validate the weighting hypothesis, a systematic double-counting project was initiated at the Specola Solare Ticinese observatory in Locarno which, as a former auxiliary station of the Zürich Observatory since 1957, continues nowadays to use this weighted counting method, providing a living memory of how Zürich observers worked back in 1945 (Cortesi et al., 2016, Ramelli et al., 2018). Based on simultaneous weighted and normal counts (Wolf's original formula), Svalgaard, Cagnotti and Cortesi (2017) found that the weighting method produced a mean inflation of about 17% on average over the studied interval (2012-2014). The amplitude of this effect thus closely matches the magnitude of the 1947 jump, thus giving a strong indication that the introduction of this weighting led to an overestimate of the $S_N(1.0)$ over the whole period after 1945, up to the present, as the Specola Observatory kept the role of pilot station after the move to the sunspot production from Zürich to Brussels in 1981.

However, the magnitude of the 1947 jump, initially estimated at 20% (Svalgaard 2012, 2013), was quickly questioned. Lockwood, Owens, and Barnard (2014a) compared the mean ratios between the $S_N(1.0)$ and independent series (Greenwich sunspot areas and group counts, inter-diurnal variability (IDV) geomagnetic index) and found a much smaller jump amplitude of 11.5% between the mean ratios over two long intervals, 1875-1945 and 1946-2012. This discrepancy was clarified by Clette and Lefèvre (2016). First, uncorrected inhomogeneities were present before 1900 in the original Greenwich group counts used as one of the references, and the choice of 1945 as the separating year (time of a cycle minimum) did not match the actual time of the jump present in the series. By just correcting those two flaws, the same analysis leads to a larger jump of about 16%. Moreover, by a finer analysis of the double counts conducted at the Specola Observatory, Clette and Lefèvre (2016) and subsequently Svalgaard, Cagnotti and Cortesi (2017) found that the inflation factor varies with the level of solar activity, starting near 1 at low activity, then increasing to an asymptotic plateau that is reached around $S_N$= 50. Above this limit, the inflation factor levels out near a mean value of 1.177. A slight upward dependency may persist for very high $S_N$, which could explain even larger values closer to 1.2, as found by Svalgaard (2013), for the maxima of very strong solar cycles, considering that this analysis was applied to cycle 24, which was weaker by a factor two than the cycles of the mid-20$^{th}$ century. Likewise, the dependency of the inflation factor on solar activity, and thus the resulting variation over a range from 1 to 1.177 in the course of each solar cycle, also explains why lower mean inflation values of about 1.15 are obtained when averaging over a full solar cycle or multiple solar cycles, like in the analysis by Lockwood, Owens and Barnard (2014a). A synthesis of those elements by the ISSI Team thus allowed us to conclude that the issue of the amplitude is now largely settled. A graphic summary of the various determinations of the amplitude of the discontinuity in S$_N$ ca. 1947 is given in Figure 5.

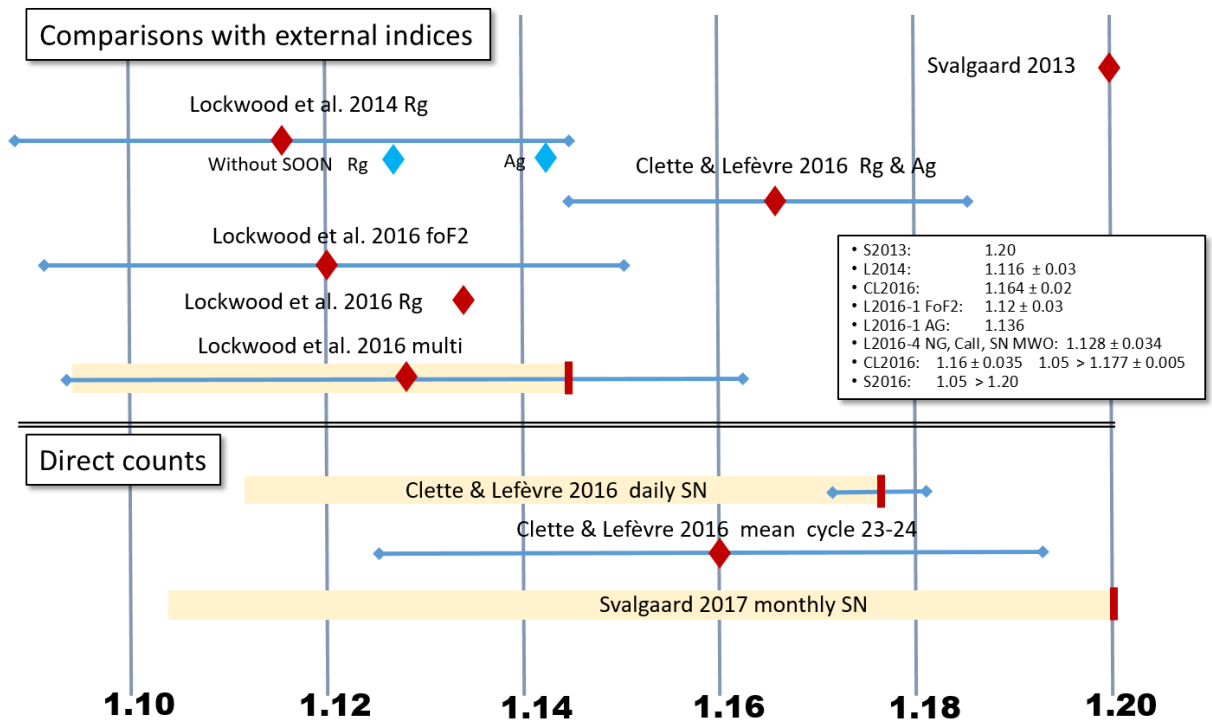

Figure 5. Summary of published determinations of the amplitude of the 1947 jump (upper part) and of the inflation factor derived from simultaneous weighted and unweighted direct counts (lower part). Red and blue diamonds indicate mean values, with the uncertainty ranges shown as blue arrows. Yellow bands indicate the range of values, when the factor is found to be variable. Vertical red bars are the upper limits, typically found near solar cycle maxima. Rg = relative group sunspot number (G$_N$) and Ag = group area.

The date of this jump also raised another issue, assuming that the Zürich weighting method is truly the primary cause of the jump. Indeed, different pieces of evidence indicate that the weighting method was implemented in the early 20$^{th}$ century, decades before 1947 (Cortesi et al., 2016; Svalgaard, Cagnotti and Cortesi, 2017). Mentions in the Mittheilungen and in the Zürich archives indicate that this method was introduced by Wolfer for the Zürich assistants, apparently to help them to obtain counts matching more closely his own (unweighted) counts, as the primary observer. The use of the weighting in the first half of the 20$^{th}$ century was also verified by consulting original sunspot drawings from the Zürich

Observatory. However, surprisingly, no significant inflation is found in the counts of Broger and Brunner, the main assistant and the successor of Wolfer, who both used the weighting, except for low $S_N$ numbers below 25 (Svalgaard, Cagnotti, Cortesi, 2017; their figure 19). This very long delay before the putative effect of the pre-existing weighting became effective thus required an explanation. Moreover, the sharp scale jump finally occurred in 1947, as found in the data, while Waldmeier had become Director of the Zürich Observatory and primary observer in 1945, two years earlier. Therefore, this prominent change in the history of the Zürich SN construction does not even match the exact time of this anomaly. The case for a role of the weighting thus also required additional evidence of another key transition in the Zürich system.

Following the creation of the $S_N$ database and the recent recovery of all source data for the period 1945-1980, Clette et al. (2021) conducted a full survey of contributing stations since Wolf started recruiting assistants and auxiliary observers, up to the Waldmeier period that concludes the Zürich era of the $S_N$, with its extended worldwide network of auxiliary stations (cf. section 2.1.1). The resulting timelines revealed two unique disruptions in the history of the Zürich sunspot number that occurred almost simultaneously, immediately after the end of World War II. Due to the War, the former set of contributing stations was replaced by an entirely new and larger international network of auxiliary stations in the years following 1945. At the same time, in Zürich, Brunner and his primary long-term assistant continued to observe until the end of 1946, and were then replaced by Waldmeier by new assistants, among whom the first ones only stayed at the Observatory for a short time (Figure 6). Both factors produced a break in the continuity of the Zürich system, as almost all new internal and external participants never contributed before or worked in parallel with former observers. This unique event corresponds to a sudden loss of past memory of the Zürich system. This can be quantified by counting the cumulative number of past observing years for all observers active on a given year. This amount plotted as a function of time is marked by a large abrupt drop in 1947 (Figure 7), a sharp transition that is unique over the whole 1849-1980 interval. Clette et al. (2021) now show that this abrupt transition coincides with the departure of the Brunner team, which had been observing since 1926, and that it falls in 1947, precisely when the jump is diagnosed in the original $S_N$ series. This finally gives a clear historical base for the occurrence of this jump and its timing.

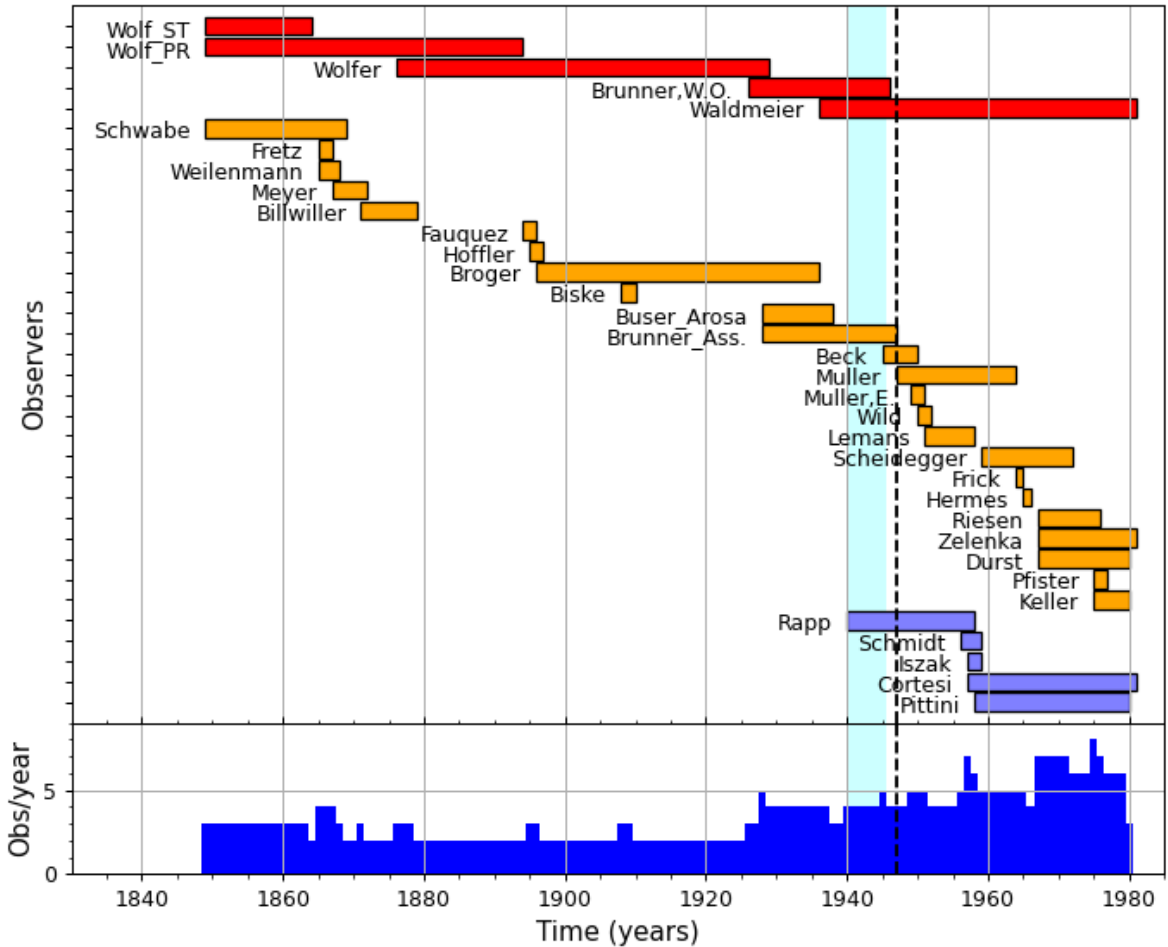

Figure 6. Timelines of the active observing periods of all Zürich observers. In red (top group), the primary observers and in orange (bottom group), the assistants. In purple, the observers of the auxiliary station in Locarno, who were considered as members of the Zürich core team. The vertical shaded band marks World War II and the vertical dashed line indicates the time when the 1947 scale jump occurs in the original $S_N$ series. The bottom plot gives the number of active Zürich observers for each year (from Clette et al. 2021).

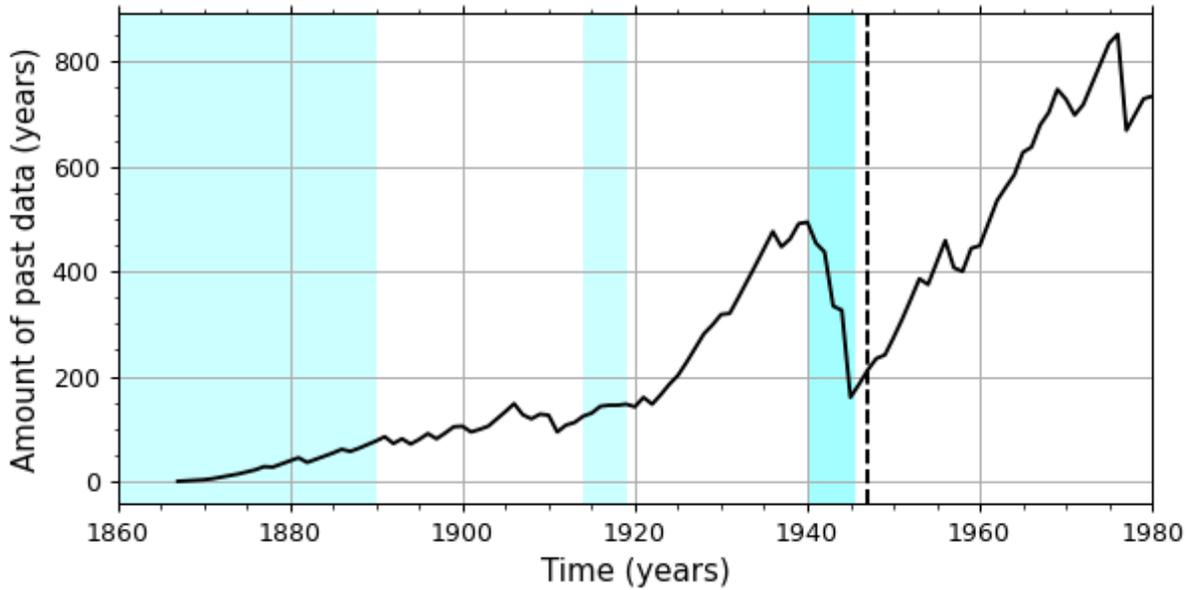

Figure 7. Plot of the total number of preceding observed years by all external stations contributing to the Zürich $S_N$ and active in each year. This count is a global measure of the amount of past information on which the $S_N$ calibration could rest for any given year. After an almost continuous increase, a sharp drop by more than a factor 2 occurred just after WWII. The shaded vertical bands mark the early Wolf period and the first and second world wars. The vertical dashed line marks the location of the 1947 jump, accompanying the last observations from Brunner (from Clette et al. 2021).

*2.1.3 Orientations for future progress*

The potential recovery of the complete Zürich set of observations, including those from 1919-1944, will make it possible to fully construct a $S_N(3.0)$ time series independently of the direct use of $S_N(1.0)$, only rescaled over certain epochs, as was done to produce $S_N(2.0)$. The methodology of constructing the $S_N(3.0)$ series has not been selected, but several options are possible, given the recent advances in methods acquired over the last few years.

Such a reconstruction could, for example, be based on the k-based method implemented at the WDC-SILSO since 1981. Non-linear probability distribution functions (Usoskin et al., 2016a; Chatzistergos et al., 2017), developed so far for the group number (described in Section 2.2.2), could also be applied after adapting them to the sunspot number. Another obvious improvement would consist in replacing the single pre-selected pilot observer by a set of reference stations, selected on the base of systematic statistical quality criteria (Clette and Lefèvre, 2016). The elements of such a method are now developed and trained on actual data from the SILSO network. This new state-of-the-art approach presented by ISSI team members (Mathieu et al. 2019, 2021), uses advanced statistical techniques to derive a non-parametric measure of the short- and long-term stability of each individual observer or observatory team, and it also introduces a proper non-normal distribution of uncertainties, in particular in the special case when the $S_N$ is close to zero, with the constraint of non-negativity. In addition to creating a new time series, such tools can form the basis for a permanent quality-monitoring process, applicable both to the past reconstruction and to the current and future production of the $S_N$ at the WDC-SILSO.

However, gathering enough base data remains the essential pre-requisite for any progress. This will thus require recovering more historical and revising the existing data series, which were often inherited indirectly through past data searches, sometimes a long time ago. For the 20th century, except for the remaining lost Zürich archives between 1919 and 1944, we now have reasonable data coverage. Still, even for this more recent period, new data series that were not known or not collected by the Zürich observatory can help completing the $S_N$ database, either with entirely new series from other observers or by extending and revising original series partly collected in Zürich. For instance, we can cite the sunspot catalog of the Madrid Observatory (Aparicio et al., 2018) and sunspot-count series from dedicated Japanese observers like Hisako Koyama (Hayakawa et al. 2020b), Hitoshi Takuma (Hayakawa et al., 2022b) or Katsue Misawa (active over 1921-1934).

As nicely illustrated by Muñoz-Jaramillo and Vaquero (2019), the main challenge resides in the sparse data of the first decades of the 19th century and in the 18th century. For the 19th century, we can incorporate recounts of Schwabe's daily sunspot numbers based on examination of his drawings (Arlt et al., 2013; R. Arlt, personal communication, 2022) to better bridge the gap corresponding to the sunspot dearth in the Dalton minimum and to shed light on the validity of the 1849-1863 "Schwabe-Wolf" scale transfer. In the mid-19th century, although Wolf produced the only long and continuous series, his method went through several meaningful changes between 1861 (introduction of k-coefficients) and 1877 (start of the Wolfer contribution), as summarized by Friedli (2016, 2020). Therefore, a full understanding of the scale stability before the Wolf-Wolfer transition described above (Section 2.1.2.2) requires a revision and extension of known data series, like the recent recounting of sunspots from the original high-quality synoptic maps by Richard Carrington over 1853-1861 (Bhattacharya et al., 2021). One important goal is to fix the scale transfer between Schwabe, the last link of historical observations, and Wolfer, the first modern long-duration sunspot observer.

In the 18th century, the data recovery effort largely merges with the one for the group number (see Section 2.2.1 below), with the extra requirement that for the $S_N$, we need the spot counts in addition to the group counts that were already collected in the $G_N$ database (Vaquero et al., 2016). As many early observers did not provide such detailed counts or often did not make the distinction between sunspots and sunspot groups, progress during this era will require consultation of original sources, in particular sunspot drawings, following the example of the recounting of Staudacher's drawings by Svalgaard (2017). Such a recounting is currently in preparation for the drawing series from De Plantade (which was considered lost; Hoyt and Schatten, 1998a,b) as a result of Hisashi Hayakawa's manuscript recovery. De Plantade's manuscript covers the first decades of the 18th century (1705-1726), and thus the critical period when the solar cycle restarted at the end of the Maunder Minimum.

2.2 Group sunspot number ($G_N$)

*2.2.1 Data recovery and revision*

For the ~400-yr span of the $G_N$ series, Muñoz-Jaramillo and Vaquero (2019) define two qualitatively different periods. The first corresponds roughly to the first two centuries of telescopic sunspot observations (1610-1825) for which the number of observations is low, and directly overlapping temporal comparisons between observers are difficult to make. The second period spans from ~1825 to the present, when the temporal coverage is better and there is a clear network of related and comparable observers.

Historical sunspot drawings are taking on an increasingly important role in space-climate studies (e.g., Muñoz-Jaramillo and Vaquero, 2019) because they constitute ground-truth raw data – with a direct picture of the shape, size, and distribution of spots and groups on the solar disc. They also give finer diagnostics to understand how the observations were made and their resulting quality. For modern times, annotated drawings indicate how group splitting was accomplished. The recovery of sunspot observations, and sunspot drawings in particular, is now essential as the current revision efforts shift from correction of existing time series to full reconstruction from raw data.

The development by Hoyt et al. (1994) and Hoyt and Schatten (1998a,b) of the group number time series was accompanied by the creation of a $G_N$ database that was revised and updated by Vaquero et al. (2016). The open-source V16 data makes it possible to apply different methodologies to compose $G_N$ time series. As a related development, within the past decade, the Historical Archive of Sunspot Observations (HASO; http://haso.unex.es/) was established at Extremadura University by José M. Vaquero (Clette et al., 2014; Cliver et al., 2015; Vaquero et al. 2016). The objective of HASO is *"to collect and preserve all documents in any format (original, photocopy, photography, microfilm, digital copy, etc.) with sunspot observations that can be used to calculate the sunspot number in the historical period or related documents."* For a comprehensive review of historical sunspot records and the recent improvements available, see Arlt and Vaquero (2020).

Since 2016, more than 30 papers have been published on sunspot data recovery with ISSI Team members as principal or co-authors. Here, we review some key results of these studies, focusing on the recovery/revision of group counts, which have been incorporated into V16. Many of the data sets that have been recently uncovered and analyzed (e.g., Carrasco et al. (2018a; for Hallaschka), Arlt (2018; for Wargentin), Nogales et al. (2020; for Oriani); (Hayakawa et al., 2021d; for Johann Christoph Müller)) provide information (counts and/or drawings) for both S and G and thus can be applied to revisions of $S_N$ as well as to $G_N$. As a manifestation of the increasing focus on recovery and archival of sunspot drawings, several recent works have constructed butterfly diagrams for historical observers (e.g., Leussu et al., 2017; Neuhäuser, Arlt, and Richter, 2018; Hayakawa et al., 2020a; Hayakawa et al., 2022a; Vokhmyanin, Arlt, and Zolotova, 2021; see Figure 9 below).

*(a) The earliest sunspot observations: 1610 – 1645*

Thomas Harriot recorded the earliest sunspot drawing that was based on telescopic observation in 1610. Vokhmyanin et al. (2020) revised Harriot's sunspot group number and reconstructed butterfly diagrams on the basis of copies of his original manuscript. Shortly afterward, in 1611, Gallilei and Scheiner started their sunspot observations. Gallilei's data quality greatly improved after Benedetto Castelli invented a method to project a solar disk on a white paper. This method update has been detected in the comparison of Gallilei's data with Harriot's data (Carrasco, Vaquero, and Gallego, 2020a). Sunspot data by Gallilei and his contemporaries have been comprehensively analyzed in Vokhmyanin et al. (2021).

Christoph Scheiner was the most active observer before the Maunder Minimum. He published his sunspot observations in `Rosa Ursina' and `Prodomus' (Scheiner, 1630, 1651). On their basis, Arlt et al. (2016) derived the sunspot positions and areas from his sunspot observations for the period 1611 – 1630. Carrasco, Vaquero, and Gallego (2022a) have studied Scheiner's sunspot group number in comparison with V16 and Arlt et al. (2016) and obtained two important results: (i) the shape of the second solar cycle of the telescopic era is similar to a standard Schwabe cycle, and (ii) the amplitude of this solar cycle (according to raw data) is significantly lower than the previous one. This last result supports a gradual transition between normal solar activity and deep solar activity in the Maunder Minimum (see, for example Vaquero, et al., 2011).

Charles Malapert was a key sunspot observer ca. 1618-1626. He is the only observer in the V16 $G_N$ database for ~60% of his 185 observation days. From an examination of Malapert's reports from 1620 and 1633, Carrasco et al. (2019a; 2019b) increased the net number of Malapert's daily observations to 251. Moreover, they determined that while Malapert sometimes drew only a single group in his drawings, he sometimes observed several groups. Therefore, Malapert's group counts, taken from the drawings, are now known to be lower limits.

Jean Tardé and Jan Smogulecz recorded sunspot observations in 1615 – 1617 and 1621 – 1625. Their results had been recorded in their books for sunspots as sunspot drawings and textual descriptions (Arlt and Vaquero, 2020). These records permit derivations and comparisons with other observers of sunspot group number and sunspot positions for these periods (Carrasco et al., 2021e).

Daniel Mögling was another key sunspot observer in 1626 – 1629. Hayakawa et al. (2021b) exploited his original manuscript in the Universitäts- und Landesbibliothek Darmstadt and confirmed his sunspot observations for 134 days. This study revised Mögling's sunspot group number as well as those of Hortensius and Schickard. It also derived Mögling's sunspot positions to construct a butterfly diagram. These results filled the data gap in the declining phase of Solar Cycle -12 showing the decay of the sunspot group number and equatorward migration of the reported groups.

Pierre Gassendi conducted sunspot observations in 1631-1638. Vokhmyanin et al. (2019) have analyzed his publications to revise the sunspot group number and derive sunspot positions.

Hevelius (1647) carried out the last known systematic sunspot records made by any astronomer before the Maunder Minimum for the period 1642 – 1645. Carrasco et al. (2019c) revised the sunspot drawings included in this documentary source as well as the textual reports, showing the good quality of the sunspot records made by Hevelius. Carrasco et al. (2019c) determined that the solar activity level calculated from the active day fraction (annual percentage of days with at least one sunspot on the Sun) just before the Maunder Minimum was significantly greater than that during the Maunder Minimum. Moreover, Carrasco et al. (2019c) confirmed Hevelius's observations of sunspot groups in both solar hemispheres in contrast with those of the Maunder Minimum that exhibited significant hemispheric asymmetry (Ribes and Nesme-Ribes, 1993).

*(b) The Maunder Minimum: 1645 – 1715*

The Maunder minimum (1645 – 1715) was an exceptional period in the recent history. Sunspot activity remained extraordinarily low during several solar cycles and the spots that did appear had a strong hemispheric asymmetry, with preference for the southern solar hemisphere (Eddy, 1976; Ribes and Nesme-Ribes, 1993; Usoskin et al., 2015; Riley et al., 2015). In recent years, a great effort has been made to improve our understanding of solar activity during this remarkable period. For example, the umbra-penumbra area ratio was computed by Carrasco et al. (2018b) for sunspots recorded during the Maunder Minimum. This ratio is similar to that calculated from modern data and, therefore, the absence of sunspots in the Maunder Minimum cannot be explained by changes in this parameter. On the other hand, comparisons with contemporary eclipse drawings have revealed that significant coronal streamers were apparently missing during the Maunder Minimum, unlike those of the modern solar cycles (Hayakawa et al., 2021c), substantiating a conjecture by Eddy (1976).

The combined observing intervals in the Hoyt and Schatten (1998a,b) $G_N$ database of Martin Fogelius and Heinrich Siverus, both from Hamburg, spanned the years 1661-1690, approximately half of the 1645-1700 core of the Maunder Minimum (Vaquero and Trigo, 2015). Even after removal from the Hoyt and Schatten database of several full years of continuous reported spotless days for Fogelius and Siverus (implausible because of local weather conditions at Hamburg), the two observers were the 13th and 7th most active observers (from 1661-1671 (with 318 observations) and 1671 – 1690 (1040), respectively) from 1610 to 1715 in V16. Hayakawa et al. (2021a) consulted Ettmüller (1693) and compared it with the correspondence of Fogelius in the Royal Society archives, leading to the following proposed changes to the V16 database: (1) a reduction of the number of active days for Fogelius from 26 to 3 and a corresponding (net) reduction for Siverus from 20 to 15; (2) conversion of the dates of several observations from the Julian to the Gregorian calendar; and (3) removal of all "ghost" spotless days (days with no explicit sunspot observations interpreted as spotless days) for both observers, representing 98-99% of their observations.

Sunspot records from the Eimmart Observatory of Nürnberg have been analyzed by Hayakawa et al. (2021d), based in part on Chiaki Kuroyanagi's archival investigations. The original logbooks from this observatory have been preserved in the Eimmart Archives in the National Library of Russia (St. Petersburg; fond 998). Analysis of these manuscripts removed ghost spotless days and reduced their daily patrol coverage in contrast with V16. Hayakawa et al. (2021d) revised the sunspot group number from the Eimmart Observatory for 78 days, that from the Altdorf Observatory for 4 days, and those of Johann Heinrich Hoffmann and Johann Bernhard Wideburg for 22 days and for 25 days, respectively. Among the Eimmart Archives, Johann Heinrich Müller's logbook recorded explicit spotless days and allowed us to derive robust active day fractions and a reliable $S_N$ for him in 1709 to confirm its significantly lower level of solar activity than in other small solar cycles such as cycles 14 and 24 and even those of the Dalton Minimum (Carrasco et al., 2018a; Hayakawa et al., 2020a; Carrasco et al., 2021c). These records also allowed Hayakawa et al. (2021c) to derive sunspot positions and confirmed significant hemispheric asymmetry in the southern solar hemisphere.

William Derham recorded sunspot observations at the end of the Maunder Minimum. Derham (1710) listed his observations from 1703 to 1707 in a table where he only recorded one group for each day except on 15 November 1707 when he recorded two groups. Carrasco et al. (2019a) pointed out those could not be the real number of groups observed by Derham in that period, showing a sunspot drawing made by an anonymous observer between 30 November and 2 December 1706 recording three groups. (Note that the quality of the Derham's drawings is better than the one of the anonymous drawing.) Therefore, the group counts assigned to Derham in V16 should be used with caution and Derham's counts should be revised accordingly on the basis of his original records.

Carrasco, Vaquero, and Gallego (2021b) present and analyze two sunspots recorded by Gallet in the middle of the Maunder Minimum. In addition to the sunspot observed by Gallet from 9 to 15 April 1677 (recorded by other astronomers), Gallet reported a spot group from 1 to 6 October in the same year for which there is no record of observations by others. The latitude of this sunspot was ~ 10° S, comparable to most of the sunspots observed during the second half of the Maunder Minimum.

*(c) Solar Cycles in the 18th Century: 1715 – 1795*

Solar Cycle -3 (~1711-~1723) is considered as the first solar cycle after the Maunder Minimum. Hayakawa et al. (2021e) examined the sunspot drawings of Johann Christoph Müller during this cycle to revise his sunspot group numbers (G) and derive individual sunspot numbers (S). His sunspot group numbers are significantly different from the contemporary observations of Rost. For example, on 1719 June 15, Johann Christoph Müller recorded 3 sunspot groups (Hayakawa et al., 2021e), whereas Rost's report gave 30 "sunspot groups" according to V16. Comparative analyses have revealed that Rost's data most probably described not the sunspot group number (G) but individual sunspot number (S). This manuscript also allowed Hayakawa et al. (2021e) to derive sunspot positions in both solar hemispheres. This result contrasts sunspot activities in 1719 – 1720 with those of the Maunder Minimum during which spots were predominantly in the southern hemisphere (Ribes and Nesme-Ribes, 1993; Hayakawa et al., 2021c). Another important observer, François de Plantade (1670-1741), also recorded sunspots quite systematically during the exit from the Maunder minimum, from 1705 to 1726, and will be the subject of an upcoming study.

Few sunspot records are available for the 1721-1748 interval, as shown in Figure 8 (adapted from Vaquero et al., 2016). This is the weakest link in the entire sunspot number time series. Recently, several relatively short-duration observers during this interval have been identified and documented. Johann Beyer's sunspot records in 1729 – 1730 have been examined and revised in Hayakawa et al. (2018c). Pehr Wargentin is the only observer given for 1747 in V16, with group counts reported for 17 days for which drawings are available. Arlt (2018) documented an additional 32 days with group (and individual spot) counts (but without drawings) by this observer.

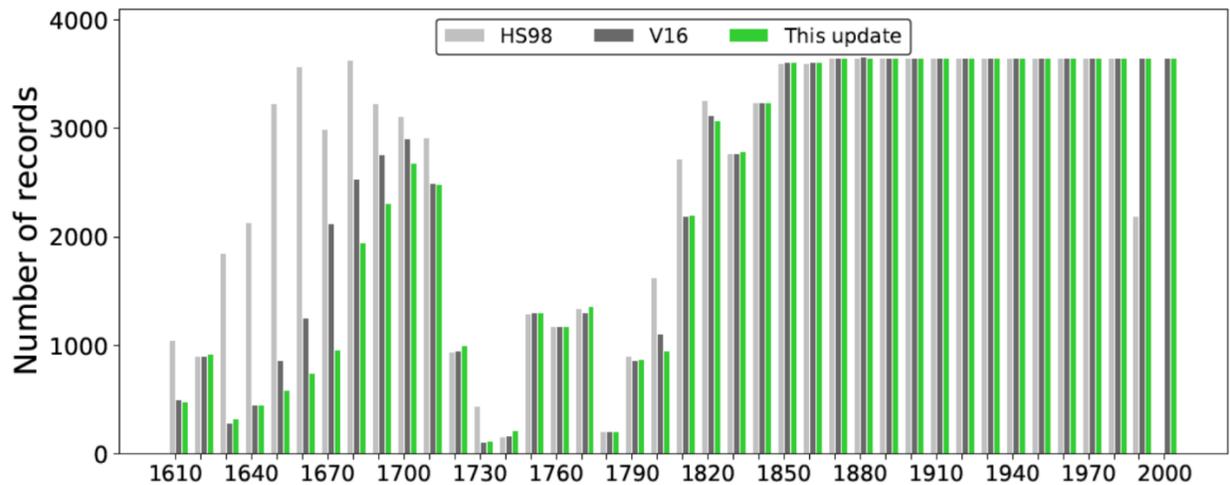

Figure 8. Number of days with records per decade in the Hoyt and Schatten (1998a,b) (gray columns) database and in its revision by Vaquero et al. (2016) (black columns). The green columns reflect subsequent modifications to the V16 data base up to the current time. The smaller numbers of records in the V16 data base for decades before 1830 is due to the removal of spurious records of days with no sunspots from the Hoyt and Schatten (1998a,b) data base. (Adapted from Vaquero et al., 2016.)

Including those data, Hayakawa et al. (2022a) have comprehensively reviewed and revised the sunspot observations in 1727-1748. Hayakawa et al. (2022a) revised the group counts of known observers, such as Krafft in 1729 and Winthrop and Muzano in 1739-1742, and added previously unknown data, such as those of Van Coesfeld in 1728-1729, Duclos in 1736, and Martin in 1738. These results have improved the morphology of Solar Cycles −2 (~1723-1733), −1 (1733-1743), and 0 (1743-1755) confirming the existence of regular cycles from 1727-1748. Hayakawa et al. (2022a) derived sunspot positions from the contemporary records and filled the data gap of the existing butterfly diagrams during this interval, confirming the occurrence of sunspots in both solar hemispheres and their equatorward migrations over each solar cycle.

Additional data have been acquired from East Asia. Hayakawa et al. (2018a,b) reported on Japanese astronomers at Fushimi and Edo (current Tokyo) and who conducted sunspot observations for 1 day in 1793 and 15 days in 1749-1750, respectively . These data fill data gaps around the neighborhood of the "lost cycle" conjectured to have occurred during the decline of cycle 4 Solar Cycle 4 in 1784-1798 (Usoskin et al., 2009; see also Karoff et al., 2015; cf., Zolotova and Ponyavin, 2011; Owens et al., 2015) and the maximum of Solar Cycle 0 (1743-1755).

Barnaba Oriani conducted sunspot observations in 1778-1779 at the Brera Observatory (Milan, Italy). Nogales et al. (2020) uncovered 52 daily sunspot observations made by Oriani for the near-maximum year of 1779 (peak in 1778) that are not included in V16. Only three other observers reported group counts for 1779, for a combined total of 19 active days. Of the 19 days, only 8 overlapped with those of Oriani, who thus accounts for 44 of 63 active days yet known for 1779. In addition, Nogales et al. determined that Oriani's group counts should be revised upward by 80% on average for his 97 daily observations for 1778 included in V16. A total of 117 active days were observed for 1878 and on 91 of these days, Oriani supplied the only observation.

Christian Horrebow's original logbooks of sunspot records are located in the Aarhus University. These records are particularly important, as his team observed sunspots from 1761-1776 and anticipated Schwabe's discovery of the sunspot cycle (Jørgensen et al., 2019). Horrebow's butterfly diagram, constructed by Karoff et al. (2019), in Figure 9 (top panel) has the characteristic structure first shown by Carrington (1858) and more definitively by Maunder (1904) for later cycles. The bottom panel in Figure 9 gives the butterfly diagram constructed from Staudacher's observations for the same interval (Arlt, 2008).

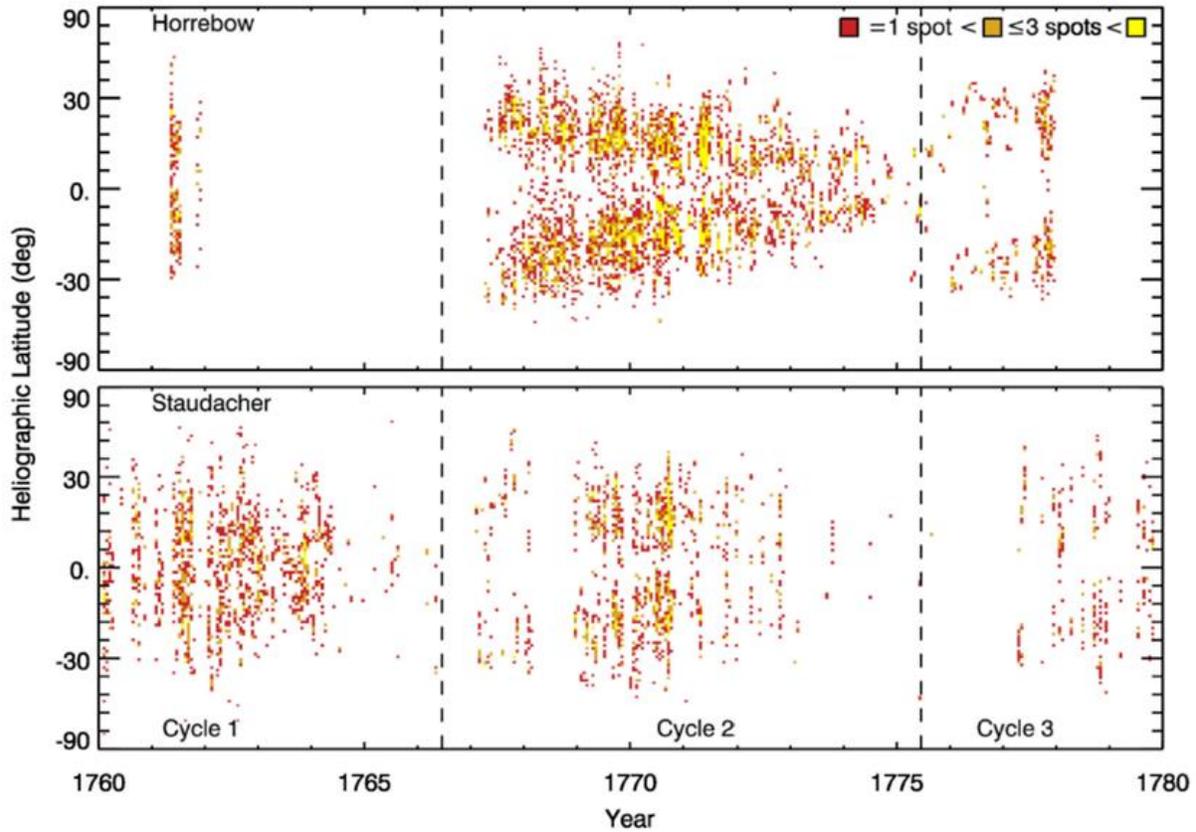

Figure 9. Top panel. Butterfly diagram constructed from Horrebow's sunspot observations Bottom panel. Butterfly diagram constructed from the observations of Staudacher, Horrebow's contemporary, for the same interval. The dashed lines indicate times of solar minima. (From Karoff et al., 2019.)

*(d) Reanalysis of a key observer improves $G_N$ records during the Dalton Minimum (1798-1833)*

The Dalton Minimum is a period of relatively low solar activity from 1798-1833, which has been named after John Dalton, who noticed a significant reduction of the auroral frequency during this time (Silverman and Hayakawa, 2021). The Dalton Minimum is similar to (though with even lower cycle peak sunspot numbers) sunspot minima that began ca. 1900 and 2010 (Feynman and Ruzmaikin, 2011). These three secular ebbs of sunspot activity, termed centennial or Gleissberg variations (after Gleissberg, 1965), are punctuated by longer periods of enhanced activity centered near ~1855 and ~1970 (Figure 2). These secular ebbs and flows of sunspot activity are less marked than the severe sunspot drought of the Maunder Minimum (Usoskin et al., 2015) and the prolonged sequence of strong solar cycles from ~1945-2008 known as the Modern Grand Maximum (Solanki et al., 2004; Clette et al., 2014; Usoskin, 2017).

The sunspot observations from 1802 to 1824 of Thaddäus Derfflinger, Director of the Kremsmünster Observatory, span the deepest part of the Dalton Minimum. From analysis of Derfflinger's drawings and associated metadata, Hayakawa et al. (2020a) concluded that the spot drawings were a secondary and therefore optional aspect of measurements of the solar elevation angle. As a result, they eliminated observations of spotless days attributed to Derfflinger, reducing the number of his daily records from 789 to 487. In addition, the butterfly diagram (Carrington, 1858; Spörer, 1880; Maunder, 1904) showing the latitudinal variation of sunspots over the solar cycle constructed by Hayakawa et al. from Derfflinger's observations was more or less symmetric about the solar equator during this period, in contrast to the deep Maunder Minimum, where spot formation occurred primarily in the southern hemisphere (Ribes and Nesme-Ribes, 1993). These results have been confirmed with Stephan Prantner's sunspot drawings for 1804-1844 from Wilten Monastery (Hayakawa et al., 2021f). Sunspots occurred preferentially in the northern hemisphere right before the Dalton minimum (Usoskin et al., 2009). These results require further investigations on the Dalton Minimum and the hypothesized 'lost cycle' at its beginning (Usoskin et al., 2009; Karoff et al., 2015; cf., Krivova et al. 2002; Zolotova and Ponyavin, 2011; Owens et al., 2015).

*(e) Solar Cycles in the 19th century*

Starting with the 19th century and even more in the 20th century, sunspot data are typically providing clearly separated counts of groups and individual spots. Therefore, those data are as relevant to the $S_N$ as to the $G_N$ reconstruction, and all the following data sets are thus contributing to the $S_N$ database described in Section 2.1.1.

After the end of the Dalton Minimum, Toubei Kunitomo conducted sunspot observations for 157 days in 1835 – 1836. While his sunspot records had been known in the existing datasets (Hoyt and Schatten, 1998a,b; Vaquero et al., 2016), preliminary analyses on the original documents have uncovered 17 additional days with Kunitomo observations in 1835 and 1 such day in 1836. Kunitomo's sunspot group number has been revised and area has been measured by Fujiyama et al. (2019). These records have filled a data gap in the existing datasets and are consistent with other sunspot observers' data, such as those of Schwabe (Fujiyama et al., 2019).

New records recovered from Antonio Colla, a meteorologist and astronomer at the Meteorological Observatory of the Parma University (Italy). Colla's records cover the period from 1830 to 1843, just after the Dalton Minimum (Carrasco et al., 2020b). Colla recorded a similar number of sunspot groups as his contemporary sunspot observers regarding common observation days. However, as is the case for Hallaschka, sunspot positions and areas recorded by Colla seem unrealistic and should not be used for scientific purposes.

William Cranch Bond was the director of the Harvard College Observatory in the mid-19th century. Bond recorded sunspot drawings from 1847 to 1849. According to V16, Bond is the observer with the highest daily number of sunspot groups observed in Solar Cycle 9 (18 groups on 26 December 1848). However, Carrasco et al. (2020c) detected mistakes in these counts. These errors are due to the use of sunspot position tables instead of the solar drawings. This new revision indicates that solar activity for Solar Cycle 9 was previously overestimated according to raw data, and Schmidt would be the observer with the highest daily group number (16 groups on 14 February 1849). A comparison between sunspot observations made by Bond, Wolf, and Schwabe (using the common observation days) shows that (i) the number of groups recorded by Bond and Wolf are similar, and (ii) Schwabe recorded more groups than Bond because he was able to observe smaller groups.

Richard Carrington made sunspot observations at Redhill Observatory in the United Kingdom, which he published in the form of a catalogue (Carrington, 1863). An observer from the current WDC-SILSO network (T.H. Teague, UK) has reanalyzed his observations (Bhattacharya et al., 2021) by recounting the groups and individual sunspots from Carrington's original drawings. Bhattacharya et al. (2021) compared Carrington's own counts (Carrington, 1863, Casas and Vaquero, 2014, Lepshokov et al., 2012) with contemporary observations, Rudolf Wolf's own observations, and Carrington's tabulations both from the Mittheilungen. They conclude that Carrington's counting methods (Carrington, 1863) for the groups were comparable to modern methods but those for individual sunspots produced significant undercounts. On the other hand, Wolf's own counts and his recounting of Carrington's drawings show numbers very similar to modern methods. The key here, is that Carrington's catalogue was, in fact, a position catalogue, thus it recorded only the biggest spots and groups, while his drawings were more precise and, when counted by Wolf in the 1860s or T.H. Teague 160 years later, give results comparable to modern observations.

Angelo Secchi observed sunspots and prominences from 1871-1875. Carrasco et al. (2021d) have constructed machine-readable tables from Secchi's book "Le Soleil" (Secchi, 1875). Secchi's original drawings indicate that he had begun sunspot observations as early as 1858 (Hayakawa et al., 2019; Ermolli and Ferrucci, 2021). These results encouraged further investigations of Secchi's original notebooks containing sunspot records in the Rome Observatory and will be the focus of an upcoming study.

*(f) Modern long-term observers*

Although the number of observers increased strongly during the 20$^{th}$ century, in particular after World War II, many series have a rather short duration and some series from professional observatories suffer from inhomogeneities due to change of instruments or observers. Therefore, the recovery of new long-duration series that were never collected and exploited, or only partly so, can help to refine the stability of the most recent part of the $S_N$ and $G_N$ records (e.g., the Zürich 1947 discontinuity found by Clette et al. 2021), and connecting it seamlessly to contemporary observations.

In this regard, sunspot observations for more recent long-term institutional and individual observers not currently included in V16 continue to be processed and digitized. The Astronomical Observatory of the Coimbra University published a catalogue with sunspot observations (including G and S) from 1929 to 1941 (Lourenço et al., 2019). In addition, a dataset of sunspot drawings made at the Sacramento Peak Observatory (SPO) from 1947-2004 has been recently digitized by Carrasco et al. (2021a). This work is the first step for the publication of the complete SPO sunspot catalogue that will include information on sunspot positions and areas. Carrasco et al. (2018c) digitized this catalogue and reconstructed the corresponding total and hemispheric $S_N$ series from Coimbra data.

The published sunspot counts of Hisako Koyama (Koyama, 1985; Knipp, Liu, Hayakawa, 2017), a staff member at the Tokyo Science Museum (later renamed the National Museum of Nature and Science (NMNS)) from 1947-1985, have been used for one of the backbones of the group number reconstruction of Svalgaard and Schatten (2016). Recent surveys of the archives of the NMNS in Tsukuba have located Koyama's sunspot drawings and logbooks from 1945 to 1996 (Hayakawa et al., 2020b). Hayakawa et al. (2020b) described and analyzed a full digital database (encoded by Toshihiro Horaguchi and Takashi Nakajima) of Koyama's sunspot observations and diagnosed a previously undetected inhomogeneity in the resulting sunspot counts affecting the later part of the series, after 1983. Hayakawa et al. (2022b) have analyzed Hitoshi Takuma's sunspot drawings from 1972-2013 in the Kawaguchi Museum. Comparisons with the contemporary records have shown Takuma's observations to be one of the most stable data sets over this ~40-yr time period.

*2.2.2 Reconstruction methodologies for $G_N$*

Several $G_N$ time series have been generated since the first such series of Hoyt and Schatten (1998a,b). These series, including that of Hoyt and Schatten (1998a,b), are listed in Table 2.

Table 2. List of studies that produced a Group Sunspot Number time series

```
Ref.*  Abbrev.**    Years      Cadence  Database       Calibration Stitching
  1    HoSc98ᵃ      1610-2013  daily    HoSc98         daisy-chain, limited backbone, linear scaling
  2    SvSc16ᵃ      1610-2015  annual   V16(1.0)       daisy-chain, backbone, linear scaling
  3    ClLi16ᵃ      1841-1976  daily    HoSc98         daisy-chain, backbone, linear scaling***
  4    UEA16ᵃ       1749-1995  daily    HoSc98         active days, synthetic reference, non-linear
  5    CEA17ᵇ       1739-2010  daily    V16(1.12)      daisy-chain, backbone, non-linear
  6    WEA17ᶜ       1749-1996  monthly  HoSc98         active days, synthetic reference, non-linear
  7    UEA21ᵈ       1749-1996  Monthly  V16(1.21)      active days, synthetic reference, non-linear
  8    DuKo22ᵉ      1825-1995  daily    V16(1.21)      tied ranking, linear
```

\* References: (1) Hoyt and Schatten (1998a,b); (2) Svalgaard and Schatten (2016); (3) Cliver and Ling (2016); (4) Usoskin et al.(2016a); (5) Chatzistergos et al. (2017); (6) Willamo, Usoskin, and Kovaltsov(2017); (7) Usoskin, Kovaltsov and Kiviaho (2021); (8)Dudok de Wit and Kopp (2022)

\*\* Time series available at: ᵃ https://www.sidc.be/silso/groupnumberv3; ᵇ http://cdsarc.u-strasbg.fr/viz-bin/qcat?J/A+A/601/A109; ᶜ https://www.aanda.org/articles/aa/full_html/2017/05/aa29839-16/ T2.html; ᵈhttps://link.springer.com/article/10.1007/s11207020 -01750-9#Sec9; ᵉ Pending publication.

\*\*\* Provisional in nature and not intended as a prescription or method for reconstruction of $G_N$; based on the Hoyt and Schatten (1998a,b) $G_N$ construction method but used an adjusted RGO time series from 1874-1915 and Schmidt (scaled to adjusted RGO)for years before 1874 as primary observers

The $G_N$ series listed in Table 2 are based on four basic methods:

1. Linear Daisy Chaining: Linear scaling of successive overlapping observers (daisy-chaining) or "backbones of observers" (Hoyt and Schatten, 1998a,b; Svalgaard and Schatten, 2016; Cliver and Ling, 2016).

2. Active Day Fractions: Independent scaling of all observers relative to a perfect observer (based on the RGO catalogue). A synthesized imperfect observer is created on a daily basis by means of a quality factor or threshold ($S_S$; analogous to k' in Equation 2) determined by the fraction of days on which spots were observed (active day fraction) (Usoskin et al., 2016a; Willamo, Usoskin, and Kovaltsov, 2017; Usoskin, Kovaltsov and Kiviaho, 2021) relative to the perfect observer. Moreover, the scaling is also based on a cross-correlation matrix, i.e., the probability distribution function (PDF) between the perfect observer and the synthesized observer, giving a non-parametric conversion.

3. Probability Distribution Function: Non-linear non-parametric scaling of successive overlapping observers via a correlation matrix (Usoskin et al., 2016a), using primary (backbone) observers, on a daily basis (Chatzistergos et al., 2017).

4. Tied Ranking: Tied-ranking method based on the rank ordering (Kendall, 1945) of group counts for a given day rather than their actual values (Dudok de Wit and Kopp, 2022).

In the following subsections, we describe each of these methods in more detail, including their advantages and shortcomings, which are summarized in Table 3.

Table 3. Summary of Sunspot Time-Series Reconstruction Methods

| Method | Requires overlap of observers | Inherently computes uncertainties | Requires parametrization assumptions | Maturity regarding sunspot series |
|---|---|---|---|---|
| **Daisy Chaining** | Yes | No | Yes | High |
| **Active Day Fraction** | No | No | No | Low |
| **Probability Distribution Function** | Yes | Yes | No | Medium |
| **Tied Ranking** | Yes | Yes | No | Low |

*2.2.2.1 Linear daisy-chaining*

This method, used by Hoyt and Schatten (1998a,b) followed the normalization approach first used by Wolf to relate his observations to those of external (to Zürich) observers before 1848. Rather than using Wolf or any other Zürich director as a primary observer, Hoyt and Schatten used the photography-based Royal Greenwich Observatory (RGO) 1874-1976 series of group numbers (Willis et al., 2013a,b; Erwin et al., 2013) as their primary reference. For observers who began observing before 1874, however, they employed "daisy chains" of individual observers in their normalization scheme, sometimes multiple chains, making it difficult, if not impossible, to replicate k'-factors for such observers (Cliver and Ling, 2016; Cliver, 2017). Hoyt and Schatten (1998a,b) were the first to use "backbones" (Svalgaard and Schatten, 2016), albeit in a limited fashion, with RGO from 1874-1976, Horrebow designated as primary observer from 1730-1800 and Galileo filling this role for the earliest observers in the 17$^{th}$ century. Svalgaard and Schatten (2016) used linear scaling of contiguous "backbones" for four primary observers.

Because the possibility for error expands with the number of contiguous links in a chain, Svalgaard and Schatten (2016) reduced the number of links in daisy chains by linking separate (mostly non-overlapping) "backbones" of observers, each scaled to a single common reference observer within a limited time interval, rather than series (single or multiple) of individual observers as Hoyt and Schatten (1998a,b) did prior to 1874. Moreover, they only used visual observers as primary references, instead of the photographic RGO group numbers. Svalgaard and Schatten (2016) used correlations of yearly averages of observer and primary counts over the interval of overlap to determine their k' factors (rather than ratios of summed daily counts for common observation days with one or more groups as used by Hoyt and Schatten, 1998a,b), forcing the fits through zero for strict proportionality. Figure 10, adapted from Chatzistergos (2017), illustrates the difference between daisy-chaining and the backbone method.

Dudok de Wit and Kopp (2022) noted the following general criticisms of the daisy-chain method:

1. Subjective initial selection of backbone observers, whose choice can significantly impact the outcome, introducing bias;
2. the method does not exploit all possible periods of overlap and in this sense is not optimal;
3. the final result is affected by the order in which the records are stitched together;
4. errors accumulate monotonically at each stitch (Lockwood et al., 2016), although the method does not inherently determine time-dependent uncertainties;
5. the method cannot temporally span data gaps (applies to all methods except Usoskin et al., 2016).

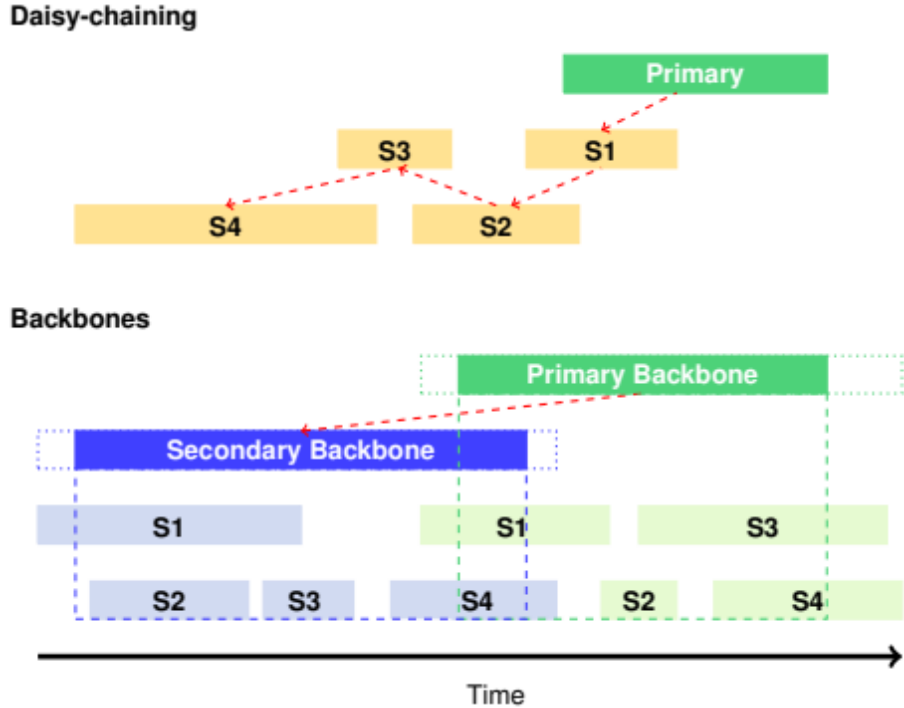

Figure 10. Schematic showing the difference between daisy-chaining and backbone methodologies. In daisy chaining, primary (green) and secondary observers (tan) are scaled sequentially to each other based on their interval of overlap. In the backbone method (Svalgaard and Schatten, 2016), several secondary observers (light-green and light-blue) are scaled to multiple backbone observers based on their interval of overlap and then the contiguous backbone series tied to each primary observer are scaled to each other. While both daisy-chaining (e.g., Hoyt and Schatten, 1998a,b) and the backbone method employ daisy-chaining, the advantage of the backbone method is the reduction of the number of links in the chain, reducing the accumulation of errors. See text for differences in the backbone approach employed by Svalgaard and Schatten (2016) and Chatzistergos et al. (2017). (Adapted from Chatzistergos, 2017.)

In the Svalgaard and Schatten (2016) backbone approach, three of the four backbone links have no overlap of the primary observers vs. four of nine such links for Chatzistergos et al. (2017). For the non-overlapping cases, the cross-calibration is based completely on the overlap of the dotted line extensions of the backbones (Figure 10). An example of non-overlapping backbone observers in the Svalgaard and Schatten (2016) reconstruction is that of Schwabe (backbone length: 1794-1883) and Wolfer (1841-1944); Schwabe ceased observing after 1867 while Wolfer's first observations (in Svalgaard and Schatten, 2016) began in 1878. The advantage of this method is that it permits long intervals of overlap with fewer backbones, but in doing so it is relying on the more uncertain parts of the extended backbone series for cross-calibration. Another difference between the backbone methodologies of Svalgaard and Schatten (2016) and Chatzistergos et al. (2017) was that secondary observers in the SvSc16 series could be scaled to both backbones, e.g., the light green S1 observer in Figure 10, but not in CEA17.

In addition, the application of a 7% reduction of the group count due to a suspected change in group-splitting technique at Zürich applied for years after 1940 in the Svalgaard and Schatten (2016) series needs to be verified independently.

A third key difference between SvSc16 and CEA17 – discussed in Section 2.2.3 – is the scaling procedure: linear scaling of yearly averages vs. a non-parametric mapping of daily values of a pair of observers. The latter allows for non-linear relations between counts of a pair of observers, and avoids side-effects of temporal averaging (different distributions of observing dates within a year, linearization effect of temporal averages). While the two series agree reasonably well after ~1880, they diverge beforehand (Figure 2).

Recently, Velasco Herrera et al. (2022) announced a reconstruction of the GN series using wavelet and machine learning techniques that supports the validity of the original HoSc98 series. However, rather than re-calibrating the $G_N$ series or exploiting the recent revised $G_N$ database (Vaquero et al. 2016), they just produce a harmonic model of the original HoSc98 series, which essentially replicates the characteristics of this input series, based on a limited set of variable periodic components. Such approaches also suffer from the stochastic nature of solar activity (e.g., Cameron and Schüssler 2019, Charbonneau 2020, Petrovay 2020). This technique may help interpolating intervals with scarce data or predominantly spotless days, on the base of periodicity assumptions, but it cannot provide diagnostics of possible flaws in the original series. On the other hand, this paper does not address any of the above issues identified in the daisy-chaining principle, as well as the homogeneity issue in the early part of the RGO photographic catalog. . For the above reasons and the fact that the Velasco Herrera et al. (2022) $G_N$ series is not a re-calibration of actual data, but rather a model, we will not further consider this series here.

*2.2.2.2 Individual non-linear scaling, based on the active day fraction, relative to a degraded perfect observer*

Usoskin et al. (2016a; see also Willamo, Usoskin, and Kovaltsov, 2017, and Usoskin, Kovaltsov and Kiviaho, 2021) introduced a normalization procedure with several novel aspects. Instead of working directly with group counts, they considered the systematic statistical relation between the relative group counts of individual observers and the number of days over which they reported spots (active days) as a fraction of the total number of days on which they observed (including spotless days), i.e., the so-called active day fraction (ADF). By using this individual indicator, the scale of each observer can then be determined independently of other observers, thus avoiding any kind of daisy chaining. A key potential advantage of this approach is that it does not rely on overlapping observations in relating any two observers, so can span large time ranges (and even gaps in observations) just as accurately as short ones. The method does rely, however, on consistency in solar activity on multi-centennial timescales because of the relatively limited duration of the RGO universal observer (1900-1976; Willamo et al., 2017).

In order to derive this scale across multiple observers, the ADF statistics must be referred to a perfect reference observer, assumed to be capable of seeing all groups down to the size of the smallest pore. For this purpose, it was assumed that the RGO photographic catalogue from 1900-1976 provided a universal reference against which any other observer could be compared regardless of whether they observed over the 1900-1976 time range of RGO data. As a scaling factor to the universal RGO observer, they determined a quality factor $S_S$ which is in fact an acuity threshold (the spot area threshold over which the individual observers starts seeing spots/groups) for each observer by matching its individual ADF statistics with the ADF obtained by synthetically degrading the universal (presumed perfect) RGO observer.

This was done by assuming that counts by imperfect observers are lowered due to their limited acuity, i.e., their inability to detect the smallest spots. This was simulated by eliminating from RGO data all sunspot groups with an area below a certain threshold $S_S$ in millionths of a solar disk (μsd). The threshold value $S_S$ that provided the best match between the thresholded RGO data and the actual ADF, A, of an observer is derived via a set of cumulative probability distribution functions (PDFs) $P(A, S_S)$ of the ADF for RGO degraded at different levels (Figure 11). Once this ADF-based $S_S$ value is determined, the correction to the group numbers themselves for the target observer is derived from the relation between the group numbers for the degraded RGO data set corresponding to this $S_S$, and the reference group number for the full RGO data set, representing the perfect observer, via their cross-probability density distribution. In this final step, instead of regressing a mathematical model, this correction was implemented in a non-parametric way, by remapping directly the raw daily group numbers for the corresponding observer, via the RGO cross-PDF, delivering the corrected group number (peak of the PDF at the given raw $G_N$ value), with uncertainties (width of the PDF at this $G_N$ value).

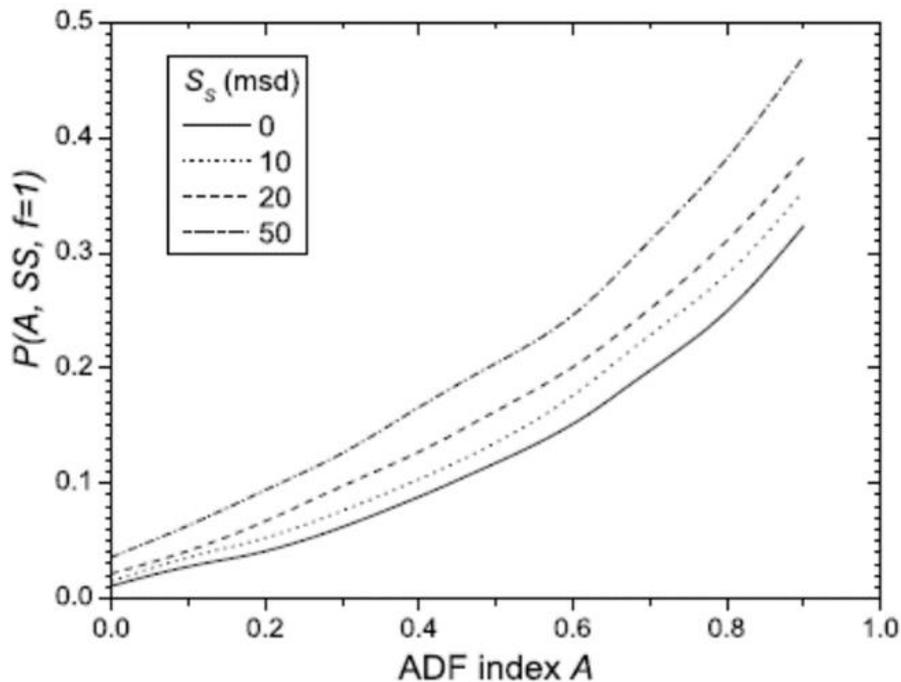

Figure 11. Cumulative probability distribution $P(A, S_S)$ for the reference data set (for different values of the threshold observed area ($S_S$) and complete data coverage (f = 1)). $S_S$ = 0 corresponds to the full RGO data set. (From Usoskin et al., 2016a.)

Difficulties/shortcomings with the ADF-based universal observer method as developed thus far include:

(1) The possibility of unreported spotless days during periods of low solar activity, resulting in an overestimation of the ADF and to underestimated corrections for some observers (Svalgaard and Schatten, 2017).

(2) Variations in the definition of a spot group (evolving group-splitting rules) used by observers from different epochs, which are not considered as another personal factor, next to the acuity of an observer. This factor can play an important or perhaps even a dominant role near solar cycle maximum, when activity is high and many groups are packed on the solar disk, and the fraction of big sunspot groups becomes large compared to tiny groups (i.e., the ones for which acuity is important). It applies to all series involving the group counts, thus both $G_N$ and $S_N$.

(3) Relative insensitivity of observer group counts to $S_S$ factors (Cliver, 2016; Svalgaard and Schatten, 2017).

(4) Applicability of the ADF approach only when the ADF is <0.8, thus only during the low part of the solar cycles. The corrected $G_N$ values for the maxima of the cycles are thus obtained by an extrapolation, outside of the range over which the ADF is calibrated for an observer.

(5) Differential sensitivity of the $S_S$ threshold to the level of activity (Usoskin, Kovaltsov, and Chatzistergos, 2016b), i.e., $S_S$ may be over- or under-estimated if the overall level of activity is different for the epoch of the target observer and for the epoch of the reference RGO data. (Solar-activity consistency is assumed.) As a result, the method works well for moderate activity but tends to slightly (<10%) underestimate high-activity levels and strongly (~30%) overestimate the low-activity levels, overall leading to a slight overestimate of the activity for the 19[th] century, given a larger occurrence of weaker cycles compared to the base time interval of the RGO data set in the 20[th] century (Willamo, Usoskin and Kovaltsov, 2018). This differential effect is probably a consequence of the extrapolation of cycle maxima mentioned in point (4) above.

(6) The observation window is different between RGO and the observer (a simple ratio of the number of days for which groups were reported to the total number of days on which observations were made is not sufficiently representative).

(7) The reference observer, in this case the RGO catalogue, is obviously not "perfect". A simple analysis reproducing the same method as in Usoskin et al. (2016a) but with different parts of the RGO catalogue (by selecting different long-term time periods) shows that the obtained $S_S$ are not consistent with the error bars reported in Usoskin et al. (2016a) (private communication, L. Lefèvre).

The above-described limitations (1-4, 6 and 7) to the applicability and accuracy of the ADF method were not addressed in Usoskin et al. (2016a), Willamo, Usoskin, and Kovaltsov (2017), or Usoskin, Kovaltsov and Kiviaho (2021). Factor (5) in the above list was quantified by Willamo, Usoskin, and Kovaltsov (2018) and Usoskin, Kovaltsov and Kiviaho (2021), but factors 1-3 may lead to larger errors and biases.

At the ISSI workshops, Muñoz-Jaramillo introduced a segmented-ADF method in order to improve the method published by Usoskin et al. (2016a). The general idea of this new method is to determine the threshold in the same way, but to compensate for points (5-7) by applying a temporal window based on the data coverage of the imperfect observer and to make it move (in time) with regard to the reference dataset to account for the level of activity within the "imperfect observer" data. Then, this threshold is applied to the reference data to count groups. Here, the actual group numbers play again a role, next to the ADF itself. Note also, that the reference dataset is slightly modified to account for possible drifts in the first years. Like the original ADF methodology for GN reconstruction, this refined segmented-ADF approach remains to be fully developed and still requires an end-to-end validation.

Willamo, Usoskin, and Kovaltsov (2017) also point out that the ADF principle is inapplicable during periods of grand minima. As is the case for daisy chaining, this approach also breaks down during intervals with sparse data. However, while the daisy-chaining methods cannot cross such data gaps, and thus reach a dead end at the first sparse-data link in the early 19[th] century, the ADF has the potential (as yet undemonstrated) to provide a calibrated tie-point to any "more populated" interval between such gaps, e.g., in the 18th century, as shown in Carrasco et al. (2021c). Thus, a hybrid approach of using the ADF method to span data gaps and daisy chaining for contiguous-observing periods is a possible strategy for creating time series over longer periods than daisy chaining alone.

*2.2.2.3 Backbones with non-linear scaling via non-parametric probability distribution functions*

Chatzistergos et al. (2017) followed Svalgaard and Schatten (2016) by reconstructing $G_N$ using a sequence of primary "backbone" observers, but improved the original concept in several respects, to avoid some of the weaknesses attributed to the initial version:

1. Use of daily values instead of yearly means.

2. Adding more primary observers, in order to have a direct temporal overlap between all of them, instead of using secondary observers to bridge temporal gaps between disconnected primary observers. (Although the probability of error accumulation increases with each link.)

3. Using a non-parametric mapping based on probability distribution functions (PDFs) of the respective values of a pair of observers, allowing for non-linear corrections, in place of least-square fitting a purely linear relation. (The non-parametric mapping method was first introduced in Usoskin et al., 2016a, as part of the ADF method described in the previous section.) Linear scaling has bias to overestimate strong cycles, while the non-parametric approach tends to overestimate minima and slightly underestimate maxima.

4. The ability to inherently estimate uncertainties in the reconstruction. In this non-parametric approach, they scaled the group counts G of secondary observers to those (G*) of primary observers through PDFs of G* for each G value. An example of a calibration matrix, for a high-quality secondary observer, Koyama, and primary observer RGO, is shown in Figure 12. This procedure makes no assumption about the type of relationship between G and G* (e.g., linearity) and the error estimate is straightforward. For each backbone, Chatzistergos et al. (2017) constructed a composite series by averaging all the PDFs of all the available observations for every day, to get a distribution based on all available observers. When there are few data points, as in the upper range of $G_N$ values, uncertainties can be estimated by applying Monte Carlo techniques to the PDFs of paired observers when creating time series.

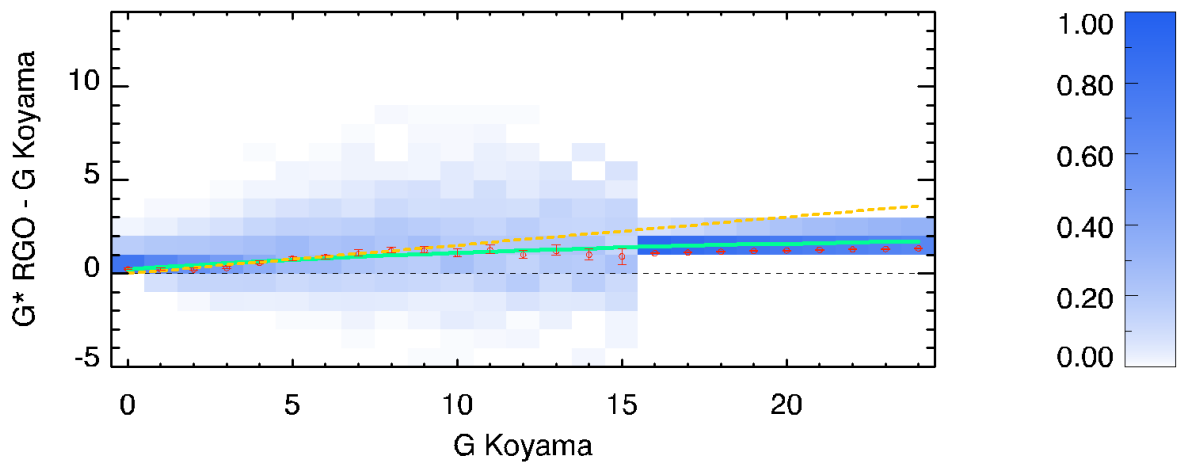

Figure 12. Calibration matrix showing the probability distribution of the residual difference between RGO (primary observer G*) and Koyama (secondary observer, G) as a function of G over 1947-1976. To compensate for the small number of data points in the upper range, columns for > 15 have been filled with the results of a Monte-Carlo simulation. The red circles with error bars depict the mean G* values for each G column and their 1σ uncertainty. The dashed red / yellow line shows the k'-factor from Hoyt and Schatten (1998a,b), and the green line is an exponential fit, showing a slight non-linearity. (See similar figures in Chatzistergos et al., 2017)

While this PDF approach using daily observations allows a more robust error analysis than the backbone-based method of Svalgaard and Schatten (see 2.2.2.1), it suffers from other limitations, including limited accuracy in the PDF for high group counts due to lower statistics than for low group counts, which is especially important to calibrate the maxima of solar cycles.

2.2.2.4 Tied ranking

Tied ranking (Kendall, 1945) is a new approach to sunspot number recalibration (Dudok de Wit and Kopp, 2022) that is not based on the GN values themselves, but instead on their distribution as measured by observer pairs. Tied ranking replaces the GN variable for a given observer by its order ("ranking") relative to all the other values of that variable. By working with ranked values rather than with original ones, one bypasses the need for correcting individual observers for their nonlinear response, which is one of the main difficulties faced by all methods in the merging process. However, ranked records can be meaningfully compared only if they span the same time interval. To fill data gaps, the expectation maximization method (Dempster, Laird, and Rubin, 1977; Rubin 1996; Little and Rubin 2002) is used. This method is a powerful generalization of the backbone and daisy chain methods in the sense that it uses all possible overlaps to fill the data gaps, avoiding the subjective choice of periods of overlap and backbones. The final composite is an average of all the available rankings on a specific day (excluding interpolated values), and then the combined ranking is turned back into group counts using a specific observer. Dudok de Wit and Kopp (2022) suggest that the absolute scale of $G_N$ series could be given by a complementary approach, e.g., ADF (Usoskin et al., 2016a).

Possible limitations of the tied ranking method include: (1) Calibration is dependent on the time interval (phase of the solar cycle, level of activity in the interval covered by the data); and (2) the method cannot account for a trend in an observer. (This latter limitation is common to all methods). In this method, several mathematical techniques are used in succession, and the way in which the output of each step may influence the subsequent ones and the final results must still be fully understood. This will require the separate analysis of intermediate steps, by creating synthetic "benchmarking" input data sets with known characteristics and imperfections. Substantial work is thus still needed for a full validation of this fully innovative approach that emerged from the work of the ISSI team.

*2.2.3 Conclusions on the reconstruction methods*

Ideally, the reconstruction problem should be separated into two parts: a scientific choice (What is the best approach for converting the different pieces of information in numbers that can be processed?), and a statistical or analytical choice (What are the best method for merging these numbers into a single composite, given their uncertainties?). The reason for decoupling the two is that the production of the composite should not influence how the raw data are interpreted and assembled into source data series.

One of the lessons learned from the ISSI team is that such a decoupling is very difficult to achieve at this stage because all these problems are so much interrelated. The general framework that is most appropriate for dealing with such problems is a probabilistic one, in which the sunspot data record from each observer is considered as a conditional distribution that depends on the different observed or unobserved parameters. These parameters may for example be the number of spots on a given day (given the resolution of the telescope, the visual acuity of the observer, etc.) or knowing that the observer did not report anything because he/she probably saw no spots during that week. The central goal then is to determine the probability to have a true sunspot number of a given value, given the various observed or unobserved parameters: p(data|parameters).

Such a probabilistic approach naturally leads to Bayesian inference [Gelman et al., 2013] which offers a natural way for estimating such probabilities, with for example a pathway for getting rid of unobserved variables by integrating them out. Bayesian thinking also offers a natural way for updating the results when new evidence comes in. Although Bayesian inference has been found to be highly effective for building composites (e.g. [Tingley et al., 2012]), there still is a long way to go before the sunspot estimation and reconstruction problem can be expressed in terms of conditional probability distributions. However, even if this is not feasible without making approximations, it forces us to express observations that are of very different types into a common and rigorous framework for which well-established methods are available.

## 3. Benchmarks for sunspot number time series constructions

Benchmarks are rules of thumb or expectations that serve as checks, or points that need to be considered, for any sunspot-based reconstruction of solar activity. They differ from proxies in that they are based solely on sunspot data.

3.1 Expectation (1): Similarity of the $S_N$ and $G_N$ time series

The on-going sunspot number recalibration effort was motivated by the expectation that the Wolf $S_N$(1.0) and Hoyt and Schatten (1998a,b) $G_N$ time series, which closely tracked each other through a broad range of solar activity during the 20$^{th}$ century, should do so throughout their common time interval, rather than abruptly diverging as they did ca. 1880 (See Figure 1 in Clette et al., 2014, and Figure 1(a) in Clette et al., 2015). This same expectation holds for the current sets of $G_N$ time series, which agree reasonably well with $S_N$ series during the 20$^{th}$ century but exhibit a broad spread in reconstructions of yearly values before ~1880 (Figure 2). At present, the most likely explanation for a separation of $G_N$ and $S_N$ series at this time lies in Wolfer's decision (when he became an assistant at Zürich in 1876) to count individual small pores as groups (see Section 2.1.2.2). The alternative – a change in the internal workings of the Sun resulting in a difference in the relative number of spots and groups at solar maxima – seems less plausible. It is clear that the years ca. 1880, based on the marked dispersion in values of the various series in Figure 2 starting near this time, present a challenge for sunspot number reconstructions.

Figure 13 shows the ratios of $S_N$(2.0) to the various $G_N$ series, after a 11-year running window smoothing, scaled to a value of 1.0 over 1920-1974. We find the $S_N/G_N$ ratio for all sunspot series to be roughly constant over the 20th century (in agreement with Svalgaard, 2020), while the various series show divergence prior to 1900. Given the broad range of activity during the ~100-year interval ~1900-2010, we would expect $S_N/G_N$ for any $G_N$ series to remain quasi-constant also before ~1900 – as is the case for the CEA17 and SvSc16 series. We note that when computing the ratios, we excluded years with $S_N < 11$. The choice of thresholds influences the computed ratios, with increasing threshold leading to higher $S_N/G_N$ ratios for all series except HoSc98, while the ratio for the series by SvSc16 gets closer to unity. We note that a non-linear relation between group and sunspot numbers has been reported, hinting at a slight dependence of the relationship between groups and sunspots to the level of activity (Clette et al., 2016).

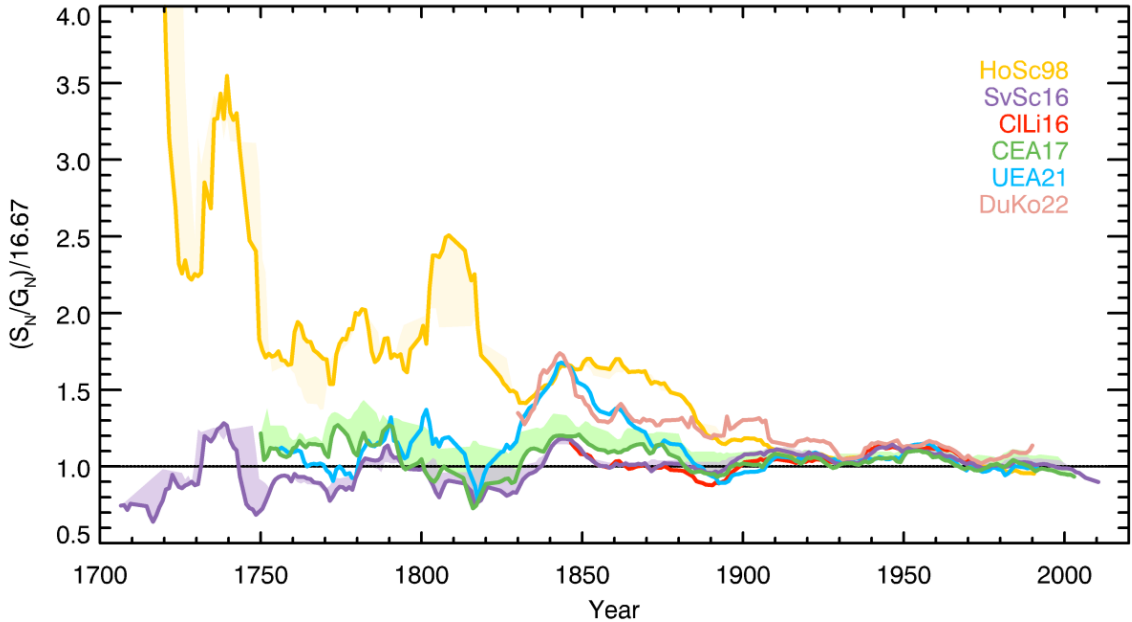

Figure 13. Ratios of the various $G_N$ series to $S_N$(2.0) after a 11-year running window smoothing, scaled to a value of 1.0 over 1920-1974. Years with $S_N < 11$ were ignored when computing the ratios. To illustrate the effect of the $S_N$ threshold on the ratios we also show as shaded surfaces (only for SvSc16, HoSc98, and CEA17) the case when years with $S_N$ less than 50 are ignored.

3.2 Expectation (2): Observers should improve, and correction factors should decrease, over time

Cliver (2016) defined a "correction factor" time series [$CF_i$] for a given $G_N$ time series (Eq. (3)), obtained by dividing the annual group count [$G_{Ni}$] by the corresponding yearly average of raw group counts for all observers [$G_{Nraw}$], that can be used to assess the reliability of new $G_N$ and $S_N$ reconstructions. $G_{Nraw}$ thus represents a fully uncorrected group number, without any compensation of the global improvement of observing techniques.

$$CF_i = G_{Ni}/G_{Nraw} \qquad (3)$$

[$G_{Nraw}$] in Cliver (2016) was produced from all observers in the V16 database and applied in Equation 3 to various $G_N$ series regardless of which data base (Hoyt and Schatten, 1998a,b or Vaquero et al., 2016) (and/or which observers within these data bases) they were constructed from. Here we produced correction factors (in essence, ensemble-averaged k- and k'-factors of observers for a given year) by considering the data used by each series and the corresponding database of raw counts. Specifically, we consulted the tables of observers listed by Cliver and Ling (2016), Chatzistergos et al. (2017), Usoskin et al. (2021a), and Dudok de Wit and Kopp (2016), and produced a [$G_{Nraw}$] series by averaging the raw counts of those specific observers from the database used in each case. For HoSc98 we used all available observers in the Hoyt and Schatten (1998a,b) database, while for SvSc16 (v1.12) and DuKo22 (v1.21), we used all observers in the V16 database, but in all cases we excluded the time series from Mt Wilson derived from just the center of the solar disk.

We would expect that in general CF values for all series would increase more or less monotonically from values ~1 in the 20th century to higher values ≳ 2 when moving back to the early 19th century and before (see Section 5.1) because of (1) inferior telescope technology for earlier centuries, and (2) the change in sunspot counting procedure from Wolf to Wolfer (see Section 2.1.2.2). Both of these changes will result in higher counts for a given level of solar activity in the modern era and a corresponding increase in CF going back in time - as can be seen for nearly all time series in Figure 14. Most GN series show to various degrees the expected decreasing trend, but it is very limited for the original HoSc98 series, which is thus incompatible with the known progress of the observations.

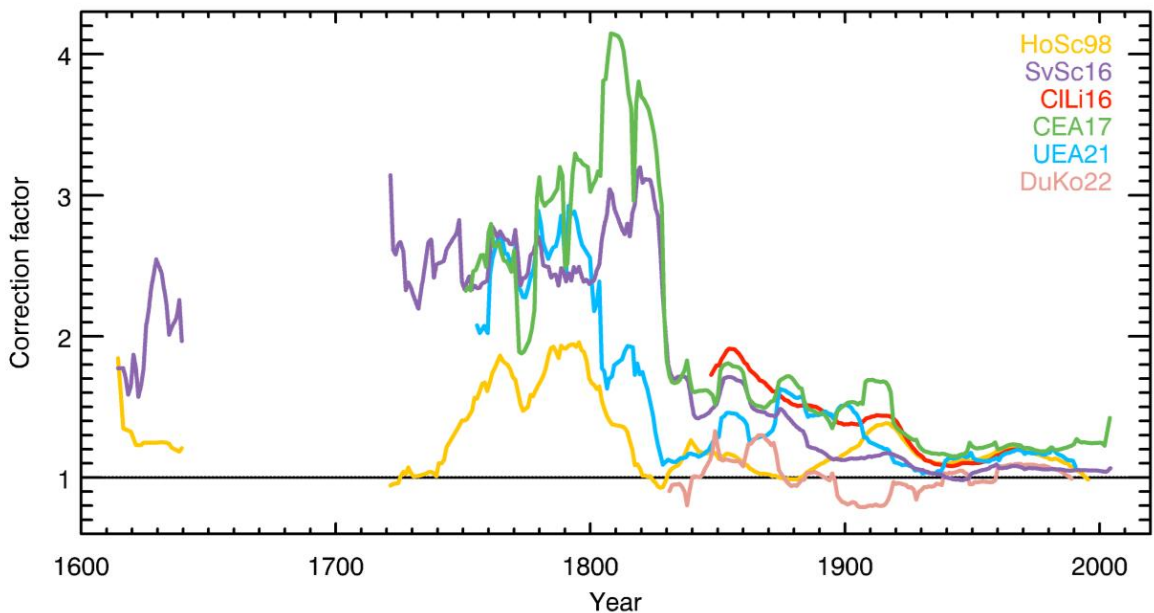

Figure 14. 11-year running averages of correction factor time series from 1600-2010 for the various sunspot series denoted in the legend.

Another kind of test can be based on determination of the losses in imperfect observations, using contemporary data. In this respect, the difference in the definitions of G and S between Wolf and Wolfer had an underlying instrumental cause that extended beyond Wolf's desire to maintain fidelity with earlier observers. After 1860, Wolf primarily used two small-aperture (40 mm/700 mm (focal length) and 42 mm/800) portable telescopes while, beginning in 1876, Wolfer used the standard 83 mm/1320 mm telescope at the Zürich Observatory (Friedli, 2016, 2020). As shown by Karachik, Pevtsov, and Nagovitsyn (2019) who degraded numerically high-resolution photospheric images from the Heliospheric and Magnetic Imager (HMI; Scherrer et al., 2012) instrument on board the Solar Dynamics Observatory (SDO; Pesnell et al., 2012) spacecraft as shown in Figure 15, telescopes with aperture < 80 mm do not resolve a significant number of small pores, and thus, likely under-estimate the group number. Telescopes with apertures larger than 80 mm, resolve the smallest pores sufficiently well, and provide a better representation of G. This is in line with the idea of quantifying the observer's quality via the acuity threshold, as discussed in Section 2.2.2.2, rather than a constant scaling factor.

Karachik, Pevtsov, and Nagovitsyn (2019) also draw attention to non-solar-induced variability of the spot-to-group ratio (see Section 3.1), writing, *"Our results indicate that there is an effect of telescope aperture on the $S_N/G_N$ ratio, which should be kept in mind while comparing modern ratios with the early observations made with small aperture instruments and using human eye as the detector."* The high values of S/G for raw (uncorrected) $G_N$ series during the 18th century are due to the inability to see small groups, i.e., groups of one or two small spots (see Section 5), due to small (<80 mm) or imperfect objective lenses.

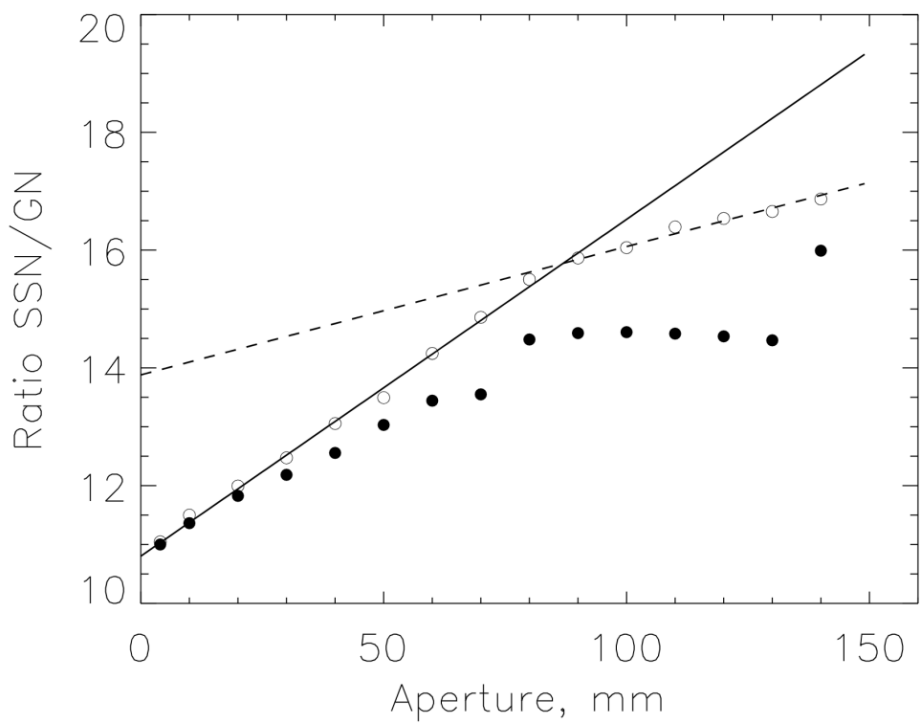

Figure 15. Dependence of the ratio $S_N/G_N$ on telescope aperture derived from numerically degraded images from HMI. Open circles – without scattered light; filled circles – with added 5% scattered light; solid line – fitted linear function for the ratio corresponding to apertures lower than 80 mm; dashed line – fitted linear function for the ratio corresponding to apertures more than 80 mm. The step-wise change in the solid circles from 130-140 mm is an artifact related to the clear aperture of 140 mm for SDO/HMI. (Based on Karachik, Pevtsov, and Nagovitsyn, 2019.)

### 4. Proxies: Independent long-term time series as cross-checks on $S_N$ and $G_N$

Different methods of sunspot number calibration include complex assumptions which may or may not be correct, and it is important to compare them to other measures of solar activity, either direct (such as solar radio emission, e.g., F10.7, or chromospheric indices, such as Ca II plage areas) or indirect (e.g., cosmogenic nuclides and geomagnetic responses). Because of large uncertainties of $S_N$ and $G_N$ in the 19th century and earlier, the use of proxy datasets can be used to corroborate sunspot estimates in the past.

Proxy data sets provide a measure of solar magnetic variability through its effects on the terrestrial environment, viz., the ionosphere via UV solar irradiance, the magnetosphere via the solar wind, or the atmosphere via the flux of cosmic rays modulated by the interplanetary magnetic field. These proxies are not affected by sunspots themselves but they are all different manifestations of the same process of solar surface magnetic activity produced by the solar dynamo in the convection zone (Charbonneau, 2020). We would expect the physical relationships of the sunspot number to such parameters to be relatively constant over time (Svalgaard, 2016; Cliver, 2017), particularly in annual averages, which remove diurnal and seasonal variations. Recent studies, however, have shown that the true relationship may be convolutive (Preminger and Walton, 2006; Dudok de Wit et al., 2018; Yeo, Solanki, and Krivova, 2020; Krivova et al., 2021), i.e., one cannot transform one solar index/proxy into another simply by assuming an instantaneous linear (or nonlinear) relationship. Indeed, solar proxies may be a delayed, cumulative, or differential response to the primary solar input. This can be explained by the fact that solar indices based on chromospheric or coronal emission also include a large contribution from the extended decay of active regions, while sunspots are much more directly tied to the initial flux emergence. Time averages of at least three months are thus expected to improve the comparisons.

Proxy data should only be used as a last resort in the construction of sunspot number time series, viz., when methods to bridge gaps in the sunspot record based on sunspot observations are unreliable, or for intervals before 1610 for which proxy $S_N$s have been based on cosmogenic nuclide data (e.g., Usoskin et al., 2021). In general, proxy data can be used both to corroborate $S_N$ and $G_N$ time series and to raise questions about their validity, particularly when abrupt discontinuities separating two extended stable periods (jumps) are observed between sunspot number time series and those for proxy parameters. The diagnostic gains in robustness if the same offset is found in comparisons of the same sunspot data series with multiple unrelated proxies or benchmarks.

4.1 2.8 GHz solar radio emission (F10.7)

Shortly after the first reported detection of solar radio waves (Southworth, 1945; Hey, 1946), it was found that the Sun's daily background 2.8 GHz emission (labelled F10.7 for the wavelength in cm) was related to sunspot activity (Pawsey, Payne-Scott, and McCready, 1946; Covington, 1947). Covington and Medd (1954) reported that F10.7 tracked the sunspot number – a close correlation that has been examined and confirmed many times since (Figure 16), most recently by Clette (2021). This good correlation can be explained by the presence at 10.7 cm of a significant contribution from gyro-synchrotron emission arising from the lower corona above sunspots. The near 75-yr span of the carefully calibrated F10.7 record beginning from 1947 (Tapping, 2013) provides a straightforward check of $S_N$ and $G_N$ series for the modern epoch. The agreement between F10.7 and sunspot number time series vouches for the physical significance of $S_N$ and $G_N$.

Yeo, Solanki, and Krivova (2020) compared various facular indices, including the F10.7, to sunspot data and found a power law function with a finite impulse response to represent the data best. In a recent in-depth study of the relation between the sunspot number and F10.7, Clette (2021) concludes that the relation between the two indices is fully linear over the whole range of values for the raw daily values. The long-known non-linearity found in the low range for $S_N < 30$ when working with monthly or yearly averages can be fully accounted for by the combined effect of temporal averaging with the non-zero minimum F10.7 background flux for a fully quiet Sun (67 sfu) and the 0-11 jump for the first sunspot in the definition of $S_N$.

This all-quiet background F10.7 flux was, by itself, the subject of various studies leading to a range of disagreeing determinations. Clette (2021) found that this lowest value is actually a function of the duration of the spotless period, increasing from 67 sfu for the longest observed intervals (30 days) up to 74 sfu for single spotless days, thereby explaining the apparently contradictory values published previously.

Moreover, by tracking the temporal evolution of the $S_N$/F10.7 relation, this study shows that the relation was fully stable over the entire 70-year interval, except for a 10.5 % jump in 1981 (Figure 17). Several tests allowed to determine that this scale jump is due to an inhomogeneity in the F10.7 series, and that it coincides with the only major historical transition in the operational production of this radio index (unique succession between the two main scientists in charge of this index, and simultaneous transition from the original manual processing to the current computerized production).

The Clette (2021) analysis supports the homogeneity of $S_N$ version 2.0 (Clette and Lefèvre, 2016). This homogeneity is also confirmed by comparisons with individual long-term sunspot observers, including some extra observers who were not included in the 2015 compilation of the SN version 2.0 (Clette, 2021; Hayakawa et al., 2022b). Equivalent comparisons of F10.7 with the original $S_N$ version 1 series and with version 2 show that the agreement between the two series is particularly improved after 1981, i.e., for the part of $S_N$ version 2.0 that resulted from a full reconstruction. Several deviations reported earlier (Yeo, Solanki, and Krivova, 2020; Clette et al., 2021) have been largely eliminated. The larger residuals before 1981 indicate that larger errors and temporary deviations remain in the Zürich part of the series, before 1980, and that the accuracy of that part of the series could still be significantly improved in future versions.

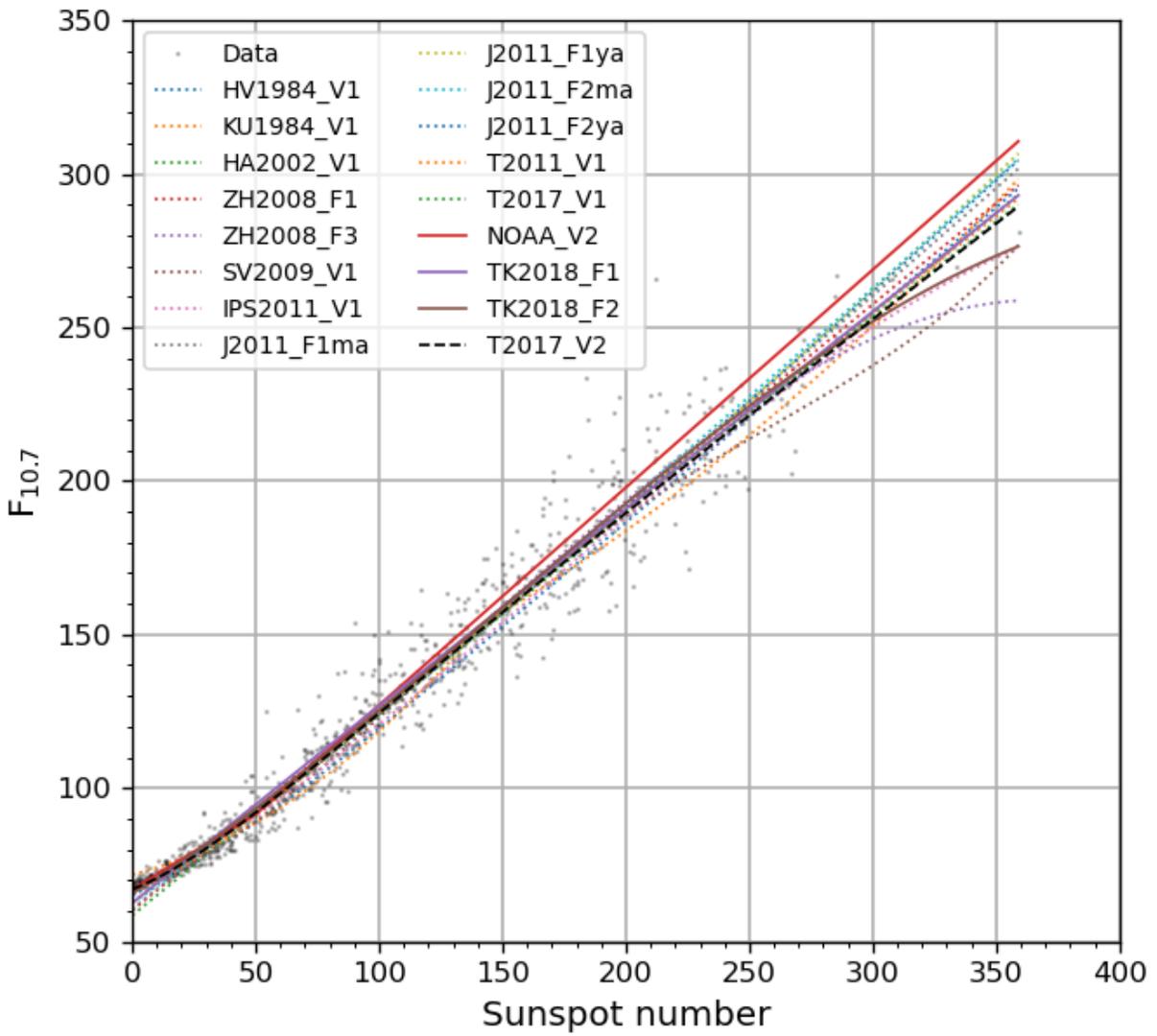

Figure 16. Plots of F10.7 vs S$_N$ (2.0) for various studies from 1984-2018. (From Clette, 2021.)

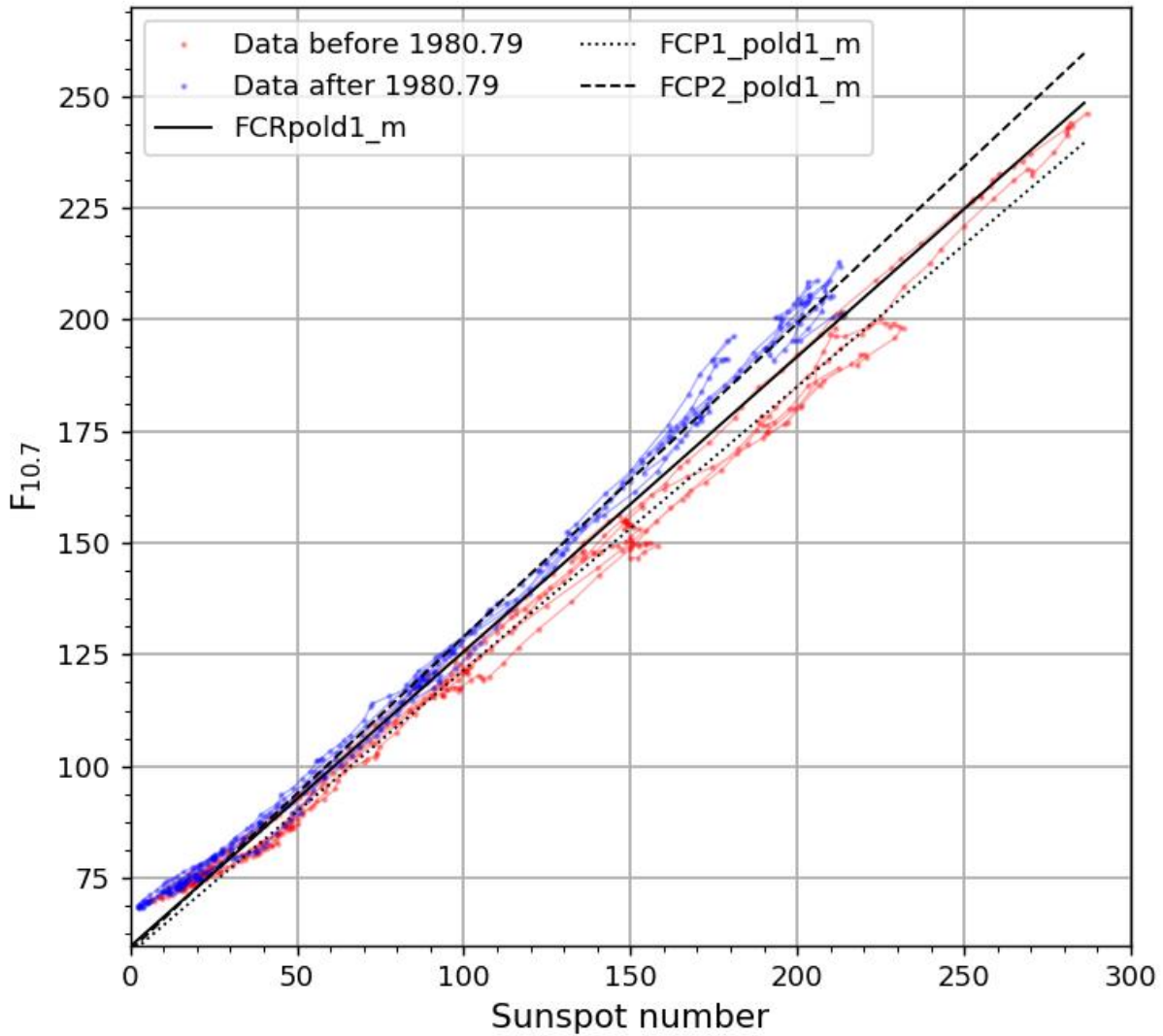

Figure 17. Plot of the 12-month smoothed monthly mean values of F10.7 versus $S_N$ (2.0) showing the largely linear relation between the two indices, and also the different slopes (F10.7/$S_N$ ratio) before 1981 (red) and after this transition (blue). The corresponding linear fits (black dotted and dashed lines) indicate a higher ratio after 1981, by 10.5%. The black line shows a global fit over the whole 70-year long series (figure from Clette, 2021).

4.2 Ca II K plage areas

Plage area series is another facular index with a connection to sunspot number series. Plage areas are determined from full-disc Ca II K (393.367 nm) observations (Chatzistergos et al. 2022b). Such observations exist since 1892 and continue to be performed from many sites around the world (Chatzistergos et al. 2022b), thus being one of the longest direct solar datasets. Various studies compared sunspot number series to plage areas (e.g., Kuriyan et al. 1982, Foukal 1996, Fligge & Solanki 1998). While more recently, Chatzistergos et al. 2022a compared plage area series from 38 archives as well as a composite series of plage areas from all available data (Chatzistergos et al. 2020) to $S_N$(1.0), SN(2.0), SvSc16, and CEA17 sunspot series. A power law relation between plage areas and sunspot number series was found to represent the data best, while a slight dependence of the relationship on the activity level was also reported. A better agreement between plage areas and $S_N(2.0)$ compared to $S_N(1.0)$ was also found, lending further support on the corrections applied to $S_N(1.0)$.

**4.3 Geomagnetic proxies**

*4.3.1 Inter-Diurnal Variation of geomagnetic activity (IDV)*

The level of energization of the Earth's magnetosphere by the near-Earth solar wind is determined by (in approximate order of importance): heliospheric magnetic field orientation and intensity, the solar wind speed, and the solar wind mass density (Vasyliunas et al., 1982; Pulkkinen, 2007). Thus geomagnetic indices, quantitative indicators of global magnetospheric disturbance based on prescribed sets of ground-based magnetometers, can be used to reconstruct near-Earth solar wind conditions (Feynman and Crooker, 1978). Given varying geometric effects associated with Earth orbit and axial inclination relative to the Sun, reconstructions are typically limited to the annual time scale, on which such effects average out (Lockwood et al., 2013). Different geomagnetic indices have different dependencies on the near-Earth solar wind conditions. Thus pairs of indices can be used to disentangle specific solar wind parameters and estimate both the near-Earth magnetic field intensity (B) and speed (V), and the open solar flux (OSF) (Svalgaard et al., 2003; Lockwood, Owens, and Barnard, 2014b). Of particular value in this regard is the inter-diurnal variation (IDV) index (Svalgaard and Cliver, 2005, 2009), which is highly correlated with B and relatively insensitive to V. Using IDV, B reconstructions can be extended back to 1845 with reasonable confidence. Fair agreement between the geomagnetic B estimates (red line) and direct in situ spacecraft observations (black) is shown in Figure 18.

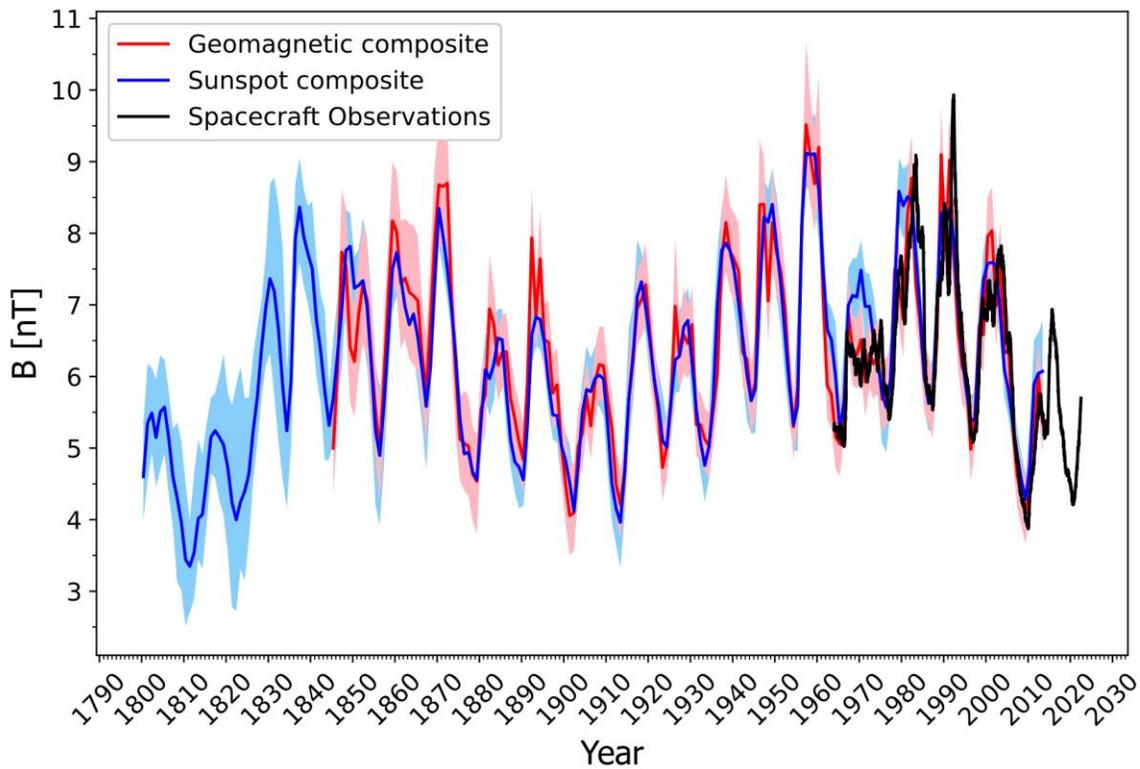

Figure 18.  Time series of near-Earth heliospheric magnetic field intensity, B, estimated from different methods. All data are annual means.  Direct observations of B, which have been made by in situ spacecraft back to 1964, are shown in black. A composite of weighted geomagnetic estimates of B are shown in red. A composite of weighted sunspot-based estimates, using a range of $S_N$ and $G_N$ time series (Clette and Lefèvre, 2016; ($S_N$(2.0)); Lockwood, Owens, and Barnard (2014a,b; LEA14);  Svalgaard and Schatten (2016; SvSc16), and Usoskin et al. (2016; UEA16), and two methods for converting the sunspot number to B, is shown in blue. The shaded regions shows the 1-sigma uncertainty ranges. (Figure adapted from Owens et al., 2016.)

In order to compare the geomagnetic estimates with $S_N$, it is necessary to convert $S_N$ to either B or open solar flux (OSF). Owens et al. (2016) considered two approaches. The first uses an empirical relationship between B and $S_N^{1/2}$ (Wang and Sheeley, 2003; Wang, Lean, and Sheeley 2005; Svalgaard and Cliver, 2005). The second uses a physically constrained model of OSF (Owens and Lockwood, 2012), which assumes sunspots are a proxy for OSF production (Solanki, Schüssler, and Fligge, 2000, 2002; Krivova et al., 2007). By assuming a constant solar wind speed, the resulting $S_N$-based OSF estimate can subsequently be converted to B.

Both methods were applied by Owens et al. (2016) to a range of $S_N$ and $G_N$ records ($S_N$(2.0), LEA14, SvSc16, UEA16). The resulting composite series is shown in blue in Figure 18. The general agreement with the geomagnetic series is strong, with the most notable deviations being an overestimate of the magnitude of B during solar cycle 20 (around 1970) and a persistent underestimate (within uncertainties) before 1900. All the individual sunspot series overestimate solar cycle 20 (1964-1976), suggesting that the difference could result from measuring global solar activity from sunspots versus the inherently local, near-Earth, measure from geomagnetic and spacecraft observations. Conversely, the higher values of B inferred from geomagnetic records before 1900 are in better agreement with the "high" $S_N$ and $G_N$ records in Figure 2 and less consistent with the original "low" sunspot records, namely HoSc98, LEA14 and UEA16.

*4.3.2 Daily range of geomagnetic activity (rY)*

The daily variation of Earth's magnetic field was first linked to the quasi-decadal variation of sunspot activity (Schwabe, 1844) in 1852 (Wolf, 1852; Gautier, 1852). During the second half of the 19$^{th}$ century, this correlation provided the strongest evidence that the magnetic field at the Earth's surface was affected by the Sun (Ellis, 1880, 1898).

Svalgaard (2016) described the physical link between the Sun's spottedness and the daily variation of the geomagnetic field as follows, *"Solar magnetism (as directly observed and as derived from its proxy the sunspot number) gives rise to an … extreme ultraviolet (EUV) excess over that expected from solar blackbody radiation …  Solar radiation into the Earth's atmosphere is controlled by the zenith angle and causes thermal winds, which, in conjunction with solar (and lunar) tides, move the atmosphere across geomagnetic-field lines.  Radiation with a short-enough wavelength ionizes atmospheric constituents (primarily molecular oxygen), and there is a balance between ion formation and subsequent rapid recombination establishing an … ionospheric conducting layer of electrons and ions [the E-layer of the ionosphere] that due to collisions moves with the winds of the neutral atmosphere across the … geomagnetic field.  The resulting inductive dynamo maintains an electric current whose magnetic effect is observable on the ground (Svalgaard, Cliver, and Le Sager, 2004; Nusinov, 2006). The day-night cycle imposes a … diurnal variation of the magnetic effect, which has been observed for several centuries. … The* *output of the entire process is the ... total daily range of the magnetic variation, which can be readily observed over a wide range of latitude."*

The annual diurnal variation, parameterized by the daily range of the (non-storm) East-component [rY] of Earth's magnetic field, has been reconstructed and tabulated by Svalgaard (2016) back to 1840 (Figure 19). It should be noted that measurements of this diurnal variation exist before 1840, but the accuracy of the resulting rY index is probably insufficient for reliable comparisons with $G_N$ and $S_N$. Indeed, as shown in Yamazaki and Maute (2017), rY has a sensitivity to seasonal effects when source data are incomplete (as is the case before 1840), because rY involves averages over whole years and over longitude.

Figure 2 shows the variation of the ratios of various scaled $S_N$ and $G_N$ series to the rY index over the interval 1840-2010. Over the 20$^{th}$ century, the gradual rise of the 11-year smoothed ratios from ~1900 to ~1975 tracks the general increase in $S_N$ and $G_N$ reconstructions during this interval, as well as the sharp drop after ~2000. This modulation of the ratio by the amplitude of the solar cycle may indicate a non-linear relation that is not fully accounted for. Nevertheless, as this relation is expected to be the same at all times, this rough agreement over 1900-2000 should hold outside of this interval. Of the nine $S_N$ and $G_N$ time series considered in the figure, CEA17 is the one for which the value of the $G_N$/rY time series is most internally consistent for corresponding extremes of solar activity in different epochs, viz., the peaks in the 18$^{th}$ and 19$^{th}$ and the troughs at the beginning of the 20$^{th}$ and 21$^{st}$ centuries. It gives a ratio that remain closest to unity. On the other hand, low reconstructions, like HoSc98 or DuKo22, strongly deviate to low ratios before 1900, thus giving the worst agreement.

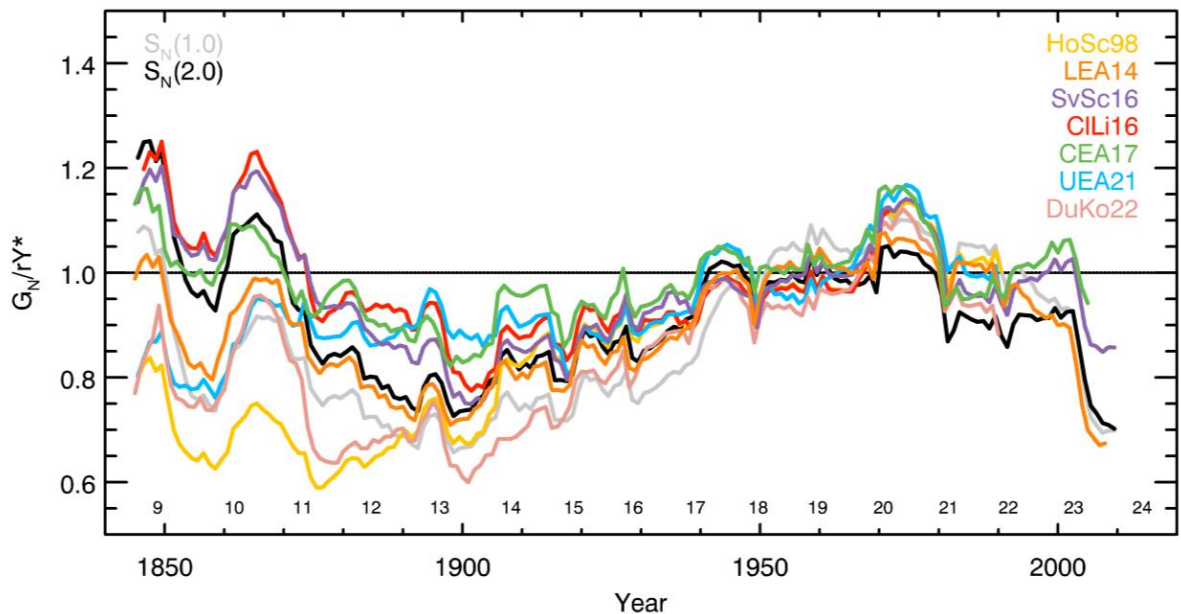

Figure 19. Ratios between the annual mean group sunspot number and annual mean range rY in nT smoothed with a 11-year running mean window. The sunspot number series were normalized to $S_N(2.0)$ over the period 1920-1974. rY was linearly scaled to CEA17 $G_N$ series to render the ratio ($G_N$/rY*) around

1. The numbers at the lower part of the panel denote the conventional solar cycle numbering and are shown roughly at the time of cycle maximum.

4.4 Cosmogenic radionuclides

Cosmogenic nuclides are radioactive isotopes which are not normally expected to exist in the terrestrial system as a result of natural radioactivity of the solid Earth, nor to survive from the time of the planetary system formation. The only (or dominant) source of such isotopes is related to energetic cosmic-ray particles continuously impinging on Earth, that initiate nucleonic-electromagnetic-muon cascade in the atmosphere. As a sub-product of the cascade, some specific radionuclides can be produced in traceable amounts and stored in natural dateable archives, such as tree trunks, ice sheets or lake/marine sediments.

The most used cosmogenic isotopes for solar-terrestrial studies are $^{14}$C (radiocarbon) and $^{10}$Be (Beer, McCracken, and von Steiger, 2012; Usoskin, 2017). The concentration of $^{14}$C in dendrochronologically dated tree rings and $^{10}$Be in glaciologically dated polar (Antarctic and Greenland) ice cores serve as measures of the abundance of these isotopes in the troposphere in the past. The flux of galactic cosmic rays near Earth is modulated by solar magnetic activity (Potgieter, 2013; Cliver, Richardson, and Ling, 2013b) which is often quantified via the modulation potential of the solar wind and heliospheric magnetic field (Caballero-Lopez and Moraal, 2004; Usoskin et al., 2005), after accounting for the effect of Earth's slowly changing geomagnetic field which provides additional shielding from cosmic rays (Usoskin, Solanki, and Korte, 2006; Snowball and Muscheler, 2007). Production tables for individual isotopes have been computed by Webber and Higbie (2003), Usoskin and Kovaltsov (2008), Webber, Higbie, and McCracken (2007), and Kovaltsov, Mishev, and Usoskin (2012). The most recent and accurate computational set was provided by Poluianov et al. (2016), which agrees well with the measurements, also in absolute terms (Asvestari and Usoskin, 2016). The first physics-based solar activity reconstruction based on cosmogenic-isotope data was made ~20 years ago (Usoskin et al., 2003; Solanki et al., 2004), and the most recent one is based on a Bayesian multi-proxy approach (Wu et al., 2018a), covering the last ten millennia of the Holocene. Such reconstructions have increasing uncertainties beyond that time because the transport and deposition patterns of isotopes in the atmosphere are less well known for ice-age conditions.

Deconvolving the cosmogenic nuclide data to infer a solar modulation potential and ultimately a proxy sunspot number for years prior to 1610 is a formidable, but necessary, task as cosmogenic isotopes provide the only quantitative information on solar activity before the telescopic era (Beer, McCracken, and von Steiger, 2012; Usoskin, 2017). One of the goals of the sunspot number workshops (Cliver, Clette, and Svalgaard, 2013a; Cliver et al., 2015) that initiated the present sunspot number reconstruction effort was to provide a robust ~400-yr sunspot series that could set the level of a cosmogenic-based sunspot number for the 10 millennia preceding 1610. As Cliver et al. (2015) wrote: *"Calibration of such a time series is complex, however, owing to variations of cosmogenic-nuclide concentrations caused by Earth's magnetic field, terrestrial climate, and possibly volcanic activity [as well as the high noise level of the raw data]. Thus it is necessary to have as long and as accurate a record of solar activity as possible to characterize the effect of these other variables on a long-term cosmogenic-nuclide-based [sunspot number]."* While the cosmogenic record works reasonably well for characterizing the relative overall level of solar activity and can distinguish between features such as the Modern Grand Maximum (Usoskin et al., 2003; Solanki et al., 2004; Usoskin, 2017) and the Maunder and Spörer Grand Minima (Eddy, 1976;

Usoskin, Solanki, and Korte, 2006; Usoskin, 2017; Asvestari et al., 2017), its use as a reliable arbiter of the smaller differences that characterize discontinuities between newly-developed $S_N$ and $G_N$ times remains to be demonstrated.

Until recently, the quality/time-resolution of the cosmogenic-isotope data was insufficient to reliably reconstruct sunspot cycles before 1600 AD (Muscheler et al., 2016). In 2021, Brehm et al. (2021) reported high-precision measurements of $^{14}C$ concentration in an oak tree archive for years after 970 AD. This new dataset together with an improved semi-empirical model describing the evolution of the Sun's global total and open magnetic flux (Krivova et al., 2021) made the first high-resolution millennium-long (970-1900) sunspot-cycle reconstruction possible (Usoskin et al., 2021b). Figure 20 compares the $^{14}C$-based sunspot number of Usoskin et al. (2021b) from 1610-1900 (gray shading indicates 67% confidence intervals) with the $S_N(2.0)$, HoSc98, SvSc16, CEA17, and UEA21 time series. Discrepancies between the amplitudes and timings of the sunspot and $^{14}C$-based time series increase as one goes back in time. Note the high minimum in the $^{14}CS_N$ record ca. 1780. The properties of individual solar cycles cannot be reliably established by this method during grand minima of activity (Usoskin et al., 2021b), but the averaged $S_N$ level is consistent with zero, implying very low $S_N$ during the Maunder minimum (see also Carrasco et al., 2021c).

Another cosmogenic-nuclide that can be used to trace the evolution of long-term solar activity is the $^{44}Ti$ isotope measured in meteorites that have fallen through the ages (Taricco et al., 2006). It is less precise than the terrestrial isotopes and can only indicate a relatively high cosmic-ray level (respectively, low solar activity) during the Maunder minimum (Asvestari et al., 2017).

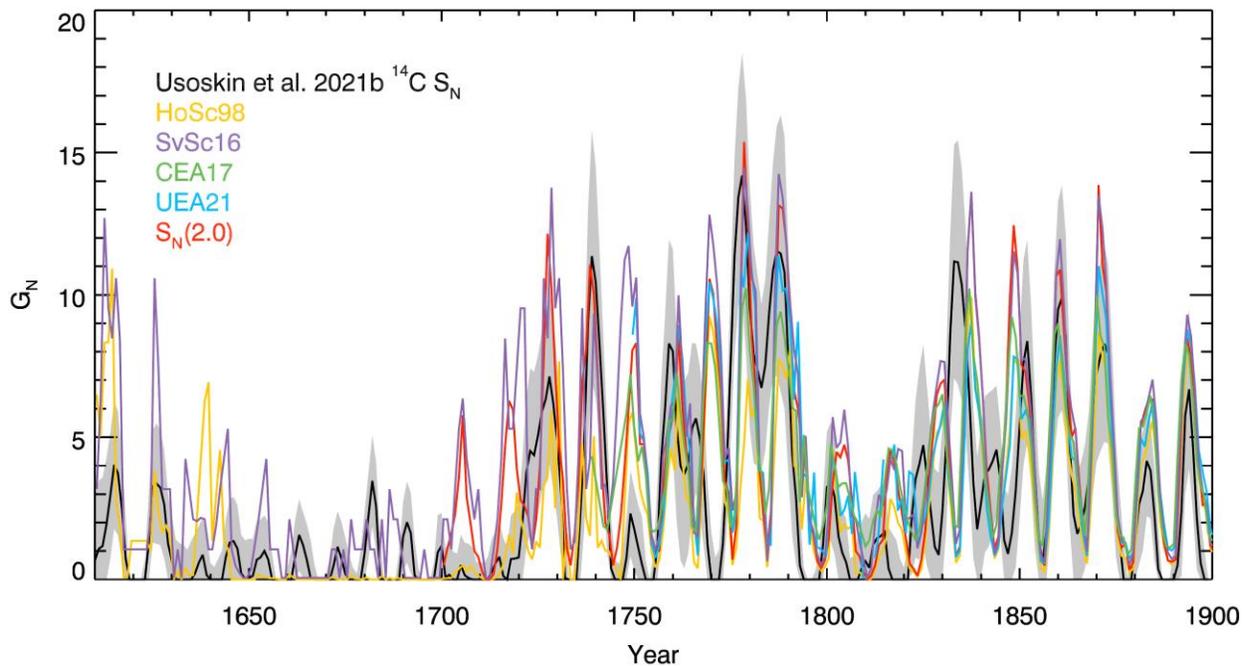

Figure 20. Comparison of $^{14}C$-based sunspot number (black curve with ±1σ grey-shaded uncertainties) with $S_N(2.0)$ (red), HoSc98 (yellow), SvSc16 (purple), CEA17 (green), and UEA21 (light blue) time series for the 1610-1900 interval. Shown are annual values, while $S_N(2.0)$ and $^{14}C$-based sunspot number were divided by 16.67 to bring them roughly to the same scale as the GSN series. All series, except the $^{14}$C-based sunspot number, were scaled to match over the period 1920-1974. (Adapted from Usoskin et al., 2021b.)

The diagnostics described above can be summarized as follows, regarding the amplitude of the reconstructed solar cycles in the 19$^{th}$ century (Table 4). Most of those tests, except the cosmogenic radionuclides, fully exclude the lowest reconstructions, which includes the original $G_N$ series (HoSc98). The intermediate reconstructions are the only ones compatible with most of those external criteria, although the agreement is never optimal (either on the lower limit or upper limit of the acceptable range).

Table 4. Summary of the diagnostics derived from the benchmarks and proxies described in Sections 3 and 4. The three categories (horizontally: low, medium, high) globally divide the total range of $G_N$ and $S_N$ reconstructions over the 19$^{th}$ century in three equal bins. The second column lists the reconstructed series that fall in those broad categories ($S_N$ series in italics). "BEST" indicates the closest match, while "NO" corresponds to a full incompatibility. When there is a "FAIR" agreement, the reconstructions are compatible with the corresponding proxy (columns) within the uncertainties, but are generally either too low or too high relative to the proxy, as indicated between brackets.

|  | $G_N$ and $S_N$ series | Observer correction factor (Fig. 14) | $S_N$(2.0)/$G_N$ (Fig. 13) | Geomagnetic IDV index (Fig. 18) | Diurnal rY modulation (Fig. 19) | Isotopes $^{14}$C, $^{44}$Ti (Fig. 20) |
|---|---|---|---|---|---|---|
| **HIGH** | SvSc16 ClLi16 *$S_N$ V2.0* | BEST | BEST | BEST | BEST | NO |
| **MEDIUM** | CEA17 DuKo22 UEA21 *SN V1.0* LEA14 | FAIR (low) | FAIR (low) | FAIR (low) | FAIR (low) | FAIR (high) |
| **LOW** | HoSc98 | NO | NO | NO | NO | BEST |

Therefore, the overall answer provided by those comparisons remains partly ambiguous. As all those external tests and proxies do not fully agree yet among themselves, further progress is thus definitely needed to improve those indirect solar tracers of solar activity, and to elucidate the remaining disagreements, before they can deliver a fully robust and independent benchmark for the $S_N$ and $G_N$ series.

## 5. Fault lines in the $G_N$ and $S_N$ data series

### 5.1 Traversing the Dalton Minimum: Staudacher to Schwabe (1798-1833)

As Muñoz-Jaramillo and Vaquero (2019) point out, there is a marked drop-off in both the quality and quantity of sunspot data before 1825 (Figure 8, Figure 21). Bridging the data-sparse period of the Dalton Minimum to scale Staudacher, who counted spots from 1749-1799, to Schwabe (1826-1867) is a key challenge for any sunspot number series. Recently recovered datasets for the Dalton Minimum will help to address this problem (Hayakawa et al., 2020a, 2021f). In general, the relative performance/accuracy of the various reconstruction methods in sparse data environments needs to be examined/assessed (e.g., Usoskin, Mursula, and Kovaltsov, 2003).

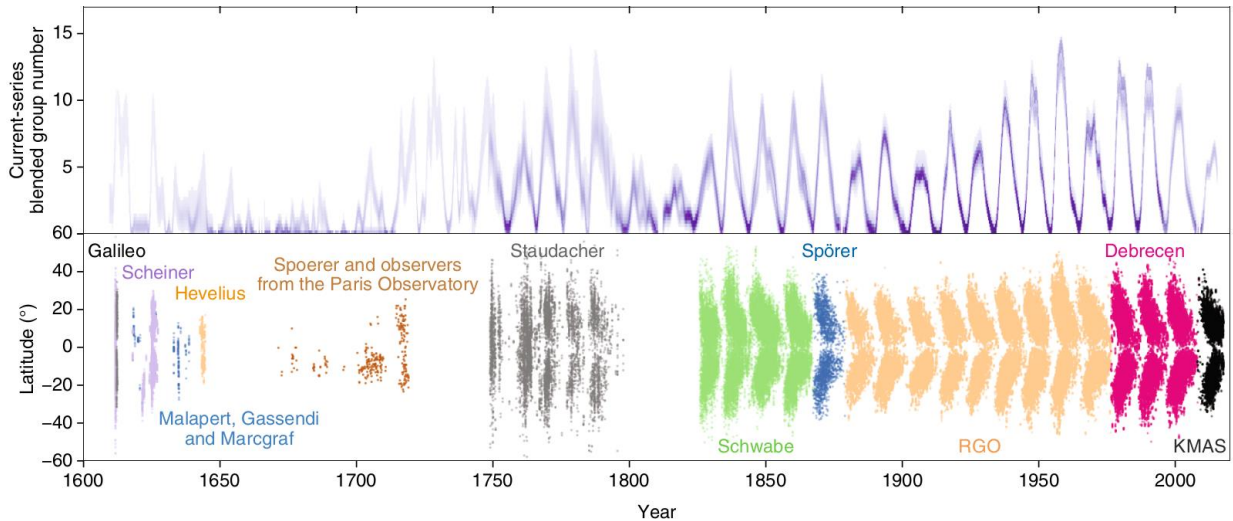

Figure 21. (Top) Composite of various $G_N$ times series. (Bottom) Composite butterfly diagram showing the separation of sunspot coverage epochs into intervals before and after Schwabe began his patrol. (Adapted from Muñoz-Jaramillo and Vaquero, 2019.)

Amateur astronomer Johann Casper Staudaucher made 1146 drawings of the spotted solar disk from 1749-1799. Quoting from Svalgaard (2020): *"Haase (1869) … reviewed the Staudach[4] material and reports that a 4-foot telescope was used, but that it was not of particular good quality and especially seemed not to have been achromatic, because he quotes Staudach himself remarking on his observation of the Venus transit in 1761 that 'for the size and color of the planet there was no sharp edge, instead it faded from the same black-brown color at the inner core to a still dark brown light red, changing into light blue, then into the high green and then to yellow'. So we may assume that the telescope suffered from spherical and chromatic aberration. We can build replicas with the same optical flaws as telescopes available and affordable to amateurs in the 18th century. On Jan. 16, 2016 we started observations of sunspots with such replicas. Three observers (expert members of "The Antique Telescope Society", http://webari.com/oldscope/) have made drawings of the solar disk by projecting the sun onto a sheet of paper. We count the number of individual spots as well as the number of groups they form. Comparing our counts with what modern observers report for the same days we find that the sunspot number calculated from the count by modern observers is three times larger as what our intrepid observers see … and that the number of groups is 2.5 times as large. This suggests that we can calibrate the 18th century observations in terms of the modern level of solar activity by using the above factors. [$S_N$(2.0)] divided by 3 … is a reasonable match to the sunspot number calculated from Staudach's drawings (Svalgaard, 2017),*

---

[4] In the literature, this observer is referred to as both Staudacher and Staudach. Here we generally refer to Staudacher (e.g., Figures 9 and 22) except when directly quoting Svalgaard.

*thus roughly validating the revised SILSO values and not compatible with the low values of the [Hoyt and Schatten 1998 (a,b)] reconstruction …."*

Svalgaard and Schatten (2016) had earlier used a factor of ~2.5 (obtained through what they described as an *"innovative (and some would say perhaps slightly dubious) analysis"* involving a comparison of "high count" and "low count" observers from ~1750 to ~1850 to scale Staudacher's observations to those of Wolfer in their $G_N$ series. The study based on antique telescopes (Svalgaard, 2020) puts this factor on firmer footing. We also note that this 2.5 value matches the correction factors found for several $G_N$ reconstructions before the 19$^{th}$ century (Figure 14).

Telescope aperture and optical quality are not the only factors that affect sunspot group count. Group splitting also plays a role. Before Hale's (1908) discovery of sunspot magnetism, closely spaced groups were generally lumped together into very large groups, sometimes spanning more than 40° in longitude, as the bipolarity of sunspot groups and the maximum possible size of an active region (~25°) were then unknown. This tendency to form overly large groups was probably also favored by the lower magnification and smaller drawing size used in early observations. In the mid-20$^{th}$ century, Waldmeier (1938) introduced a group classification system that took the temporal evolution of bipoles in a cluster of sunspots into account, abandoning proximity as the sole, or primary, criterion for identifying groups (Friedli, 2009; Svalgaard and Schatten, 2016). As a result, clusters of sunspots that previously had been counted as a single group could now be divided (split) into two or more groups (see Figure 1(b) for an example of such splitting), thus raising the group counts.

Arlt (2008; Arlt and Vaquero, 2020) suggested that Staudacher, with his 18th century telescope, missed all of the small A and B type spot groups (according to the Waldmeier classification) that make up 30-50% of all groups seen today (Clette et al., 2014). These reductions in group counts would imply a Staudacher k'-factor (Equation 2) ranging from ~1.4 to 2.0. From an analysis of Staudacher's sunspot drawings, Svalgaard (2017) found that Waldmeier's classification system would increase Staudacher's group counts in V16 by 25%. Combining the instrumental correction factor (1.43-2.0) with that for group splitting (1.25) yields a group scaling factor (k') range for Staudacher (to Wolfer) from of ~1.8 to 2.5, in good agreement with the above reconstitution using replicas of historical telescopes.

5.2 Galileo to Staudacher: Encompassing the Maunder Minimum

Figure 8 shows that the 1730s and 1740s are the weakest link in the sunspot number time series. Substantial attention has been focused on this data-poor interval (Section 2.2.1(c)) by Hayakawa et al. (2022a) which is critical to connect Staudacher to the Maunder Minimum (Section 2.2.1(b); the low end of the lever arm for TSI reconstruction and climate change studies) and all preceding years, including those for which the sunspot record must be inferred from cosmogenic radionuclides.

The degree of difficulty of getting the first ~140 years of a sunspot number series correct is underscored by the fact that there are only two systematic reconstructions of such a series that extend to 1610: (1) the Hoyt and Schatten (1998a,b) $G_N$ series that assigned an average k-factor of 1.255 (slightly higher than the average correction factor for observers during the second half of the 20$^{th}$ century; Figure 14) to 78 of 171 pre-1749 observers that did not have any common days of observation with other observers, with a median value of 1.002 for the remaining 93 observers (vs. 1.000 for RGO); and (2) the Svalgaard and Schatten (2016) $G_N$ series for which the pre-Schwabe years are based in large part on the high-count/low-count observer comparison scheme used to scale Staudacher to Schwabe (Section 5.1) and a somewhat related "brightest star" method based on the highest daily group count recorded in a given year by any observer. Such speculative methods and reliance on proxies for corroboration will likely be required to obtain yearly values from 1610-1748. The active day fraction (Kovaltsov et al. 2004, Vaquero et al., 2015, Usoskin 2017) has been used to estimate the general level of activity during the Maunder Minimum (Carrasco et al. 2021c, 2022b), which was relatively well-observed (Figure 8), and may be useful to firm up estimates elsewhere in the early series, pending the recovery of more historical data.

## 6. Summary of Progress

The main achievement of the ISSI Sunspot Number Recalibration Team was the data recovery effort headed by Arlt, Carrasco, Clette, Friedli, Hayakawa, and Vaquero, with an emphasis on the identification, digitization, and analysis of primary records, images in particular. Such data can play a key role to bridge gaps between early (pre-Schwabe) segments of the sunspot record. (Sections 2.1.1 and 2.2.1)

A prime focus of the Team meetings was the presentation/probing of novel $G_N$ reconstruction methods by Chatzistergos, Dudok de Wit, Kopp, Lefèvre, Mathieu, Muñoz-Jaramillo, Svalgaard, and Usoskin, with an emphasis on the application of modern statistical methods and determination of uncertainties (e.g., Mathieu et al., 2019) as summarized in Tables 2 and 3.

Less progress was made during the Team meetings on proxy time series, although considerable progress had been made beforehand by Svalgaard, Usoskin, Lockwood, Owens and others on long-term geomagnetic and cosmogenic-nuclide-based time series, and Clette has carried out a thorough comparison of F10.7 and $S_N(2.0)$ as a follow up of discussions in the Team meetings. These proxies can be used to corroborate as well as to identify potential weaknesses in new time series. (Section 4)

During the ISSI Team meetings, Clette introduced the concept of benchmarks for the reconstruction project, e.g., the concept that because of improvements in telescope technology and changes in the definitions of spots (reduced minimum size) and groups (from lumping to splitting), one would expect modern observers to count more spots than those preceding Wolfer (leading to smaller normalization or correction factors over time for a given level of sunspot activity), an expectation that the Hoyt and Schatten $G_N$ series failed to meet. (Section 3)

"Reverse engineering" experiments based on old (Svalgaard) and new (Karachik, Pevtsov, and Nagovitsyn, 2019) telescopes helped to assess the effect of improvements in telescope technology on the observability of sunspots over time. These studies are relevant for the scaling of Staudacher, the key observer for the second half of the 18[th] century, to Schwabe, the principal observer for the first half of the 18[th] century. (Sections 3.2 and 5.1)

At the team meetings, Pesnell represented the space forecasting community, Van Driel-Gesztelyi served as rapporteur, and Kopp reviewed the "triad" results (see below) and mapped the path forward.

Perhaps the most important marker of progress during the ISSI Team meetings was the joining of key stakeholders to argue/debate/discuss the reconstruction of the $S_N$ and $G_N$ number time series, with commitment to the goal of producing the optimal series with the data at hand and current best practice methodology. As Clette and Lefèvre (2016), wrote, the present focus on the sunspot number "marks a fundamental transition between the earlier unalterable and unquestioned data series to a genuine measurement series, like any other physical data series. As for any other measurement, it is natural to revise it as new data sources and new analysis methods become available."

## 7. Perspective and Prospect

The sunspot number has been called the longest running experiment in science (Owens, 2013). The renewed emphasis on this time series reflects the Sun's impact on our increasingly technology-based society and the need to better quantify the time-varying solar input to the terrestrial climate.

The end goal of the present effort that began in 2011 and continued through the Topical Issue in 2016 and the ISSI Team meetings of 2018 and 2019 to the present day remain the same: to produce a community-vetted series (Cliver, Clette, and Svalgaard, 2013) with quantified time-dependent uncertainties (e.g., Dudok de Wit, Lefèvre, and Clette, 2016) for the last ~400 years. Such a base reference can be used to anchor a millennial-scale cosmogenic-nuclide-based time series of solar variability dating back to the last glacial period and to test and validate physical models of the coupling between the past solar input and the observed response of the Earth system.

The last decade has shown that this process of acquiring consensus via applications and discussions of multiple approaches followed by reviews of their results cannot be rushed. The Hoyt and Schatten (1998a,b) time series, valuable both for the introduction of a $G_N$ series that could be extended to 1610 and for the creation of the first publicly available digitized database, taught the lesson of the necessity for due diligence regarding modifications/revisions of the $S_N$. The now apparent issues with the Hoyt and Schatten $G_N$ time series were not independently examined for more than a decade (Cliver and Ling, 2016), during which time it became entrenched as an alternative to $S_N$, in part because of its extension to the Maunder Minimum. In Appendix 1 of Hoyt and Schatten (1998a,b), the k'-factors for Wolf (1.117) and Wolfer (1.094) are within 2% of each other, despite the fact that Wolfer counted 65% more groups than Wolf (Svalgaard, 2013; Cliver, Clette, Svalgaard, et al., 2013a). Had this peculiarity been noticed when the Hoyt and Schatten (1998a,b) series was introduced, it is unlikely that their then new $G_N$ series would have gained the traction/usage which it retains at some level to this day (e.g., Coddington et al., 2019; Wang and Lean, 2021; Krivova et al., 2021). To ensure that all new $S_N$ and $G_N$ time series were subjected to scrutiny, the ISSI Team formation was preceded by an informal re-examination of each new sunspot number time series by separate "triads" consisting of an advocate, critic, and mediator. This format had the advantages of having former competitors working together – fostering both critical analysis and team building in anticipation of the ISSI effort.

What is then the way forward? The interactions between the members of the ISSI team have shown that the reconstruction of the sunspot number is a multifaceted and highly multidisciplinary problem. At a more conceptual level, this problem consists in collecting information of various types and origins, which are linked in different ways to the main observable of interest, which is the number of sunspots or sunspot groups. Ideally, one should decompose such a problem into two parts: a scientific choice and a statistical or analytical choice (Section 2.2.3). One of the main benefits of this exercise is that it makes us think in a probabilistic way, i.e., never separate an observation and its uncertainty.

Keeping this in mind, the expanded ISSI Sunspot Team of observers, analysts, and modelers will remain electronically connected (with the welcome prospect of actual meetings in time), with the immediate goals of refining the methodology for the various series proposed, and creating corresponding $S_N$ and $G_N$ time series with realistic uncertainties. Once the new/revised time series have been developed, they will need to be independently reproduced and evaluated using benchmarks and proxy series. Certain methods may work better in data sparse environments, so composite methodologies may be required. This will take time.

We estimate that new databases will be available/released by early 2023, with the key reconstruction methods brought to maturity and corresponding series created by the end of that year. At that point, an evaluation team chaired by the SILSO Director, will select next versions of $S_N$ and $G_N$ for sanction by an International Astronomical Union (IAU) reviewing body, with formal release targeted for conjunction with the IAU General Assembly in 2024. Even with the release of $S_N(3.0)$ and a consensus $G_N(3.0)$, it is possible that the two series may not be complete reconstructions for years before 1750, given the broader uncertainties and more complex and indirect validations required for these early years. However, values with appropriate uncertainties will be provided for these early years for both series back to 1610, to include the lever arm of the Maunder Minimum. The new series will mark the next step in a now-permanent improvement and quality-insurance process. The new series will be continuously monitored by comparison with a basket of high-quality observers, as well as with proxies and benchmarks. As new data, new knowledge, and new mathematical tools continue to emerge, on a regular basis follow-on versions will be released at intervals of 3 to 10 years, when enough material has accumulated to warrant robust and substantial modification to the series. The goal is to provide the scientific community with a unique and trusted reference that summarizes our best knowledge about the long-term evolution of the solar activity, as traced by sunspots, and a reliable link to cosmogenic nuclide data for years before 1610.

**Acknowledgments**
This work was facilitated by an International Space Science Institute (ISSI) International Team selected in 2017 as number 417, "Calibration of the Sunspot Number Series", organized by Matt Owens and Frédéric Clette. We thank ISSI for support of the team. We thank John Leibacher for encouraging us to use this status report to update the Solar Physics 2016 Topical Issue on Sunspot Number Recalibration edited by F. Clette, E.W. Cliver, L. Lefèvre, J.M. Vaquero, and L. Svalgaard.
This research has made use of NASA's Astrophysics Data System. The National Solar Observatory (NSO) is operated by the Association of Universities for Research in Astronomy (AURA), Inc., under cooperative agreement with the National Science Foundation.
F. Clette and L. Lefèvre are supported by the Belgian Solar-Terrestrial Center of Excellence (STCE) funded by the Belgian Science Policy Office. L. Lefèvre acknowledges funding from the Belgian Federal Science Policy Office (BELSPO) through the BRAIN VAL-U-SUN project (BR/165/A3/VALU-SUN).
T. Chatzistergos acknowledges support by the German Federal Ministry of Education and Research (Project No. 01LG1909C).
H. Hayakawa acknowledges financial support by JSPS Grant-in-Aids JP20K22367, JP20K20918, JP20H05643, and JP21K13957, JSPS Overseas Challenge Program for Young Researchers, and the ISEE director's leadership fund for FY2021, Young Leader Cultivation (YLC) programme of Nagoya University, Tokai Pathways to Global Excellence (Nagoya University) of the Strategic Professional Development Program for Young Researchers (MEXT), and the young researcher units for the advancement of new and undeveloped fields, Institute for Advanced Research, Nagoya University of the Program for Promoting the Enhancement of Research Universities.

H. Hayakawa thanks Chiaki Kuroyanagi for archival investigations for the Eimmart Collection in St. Petersburg, and the NIHU/NINJL Citizen Science Project for supporting his research on Misawa's sunspot records.

I. Usoskin acknowledges partial support of the Academy of Finland (project ESPERA No. 321882).

L. van Driel-Gesztelyi acknowledges the Hungarian National Research, Development and Innovation Office grant OTKA K-131508.

Usoskin, I.G., Arlt, R., Asvestari, E., et al.: 2015, The Maunder minimum (1645-1715) was indeed a grand minimum: A reassessment of multiple datasets. *Astron. Astrophys.* **581**, A95. DOI 10.1051/0004-6361/201526652

Usoskin, I.G., Kovaltsov, G.A., Lockwood, M., Mursula, K., Owens, M., Solanki, S.K.: 2016a, A new calibrated sunspot group series since 1749: statistics of active day fractions. *Solar Phys.* **291**, 2685. DOI 10.1007/s11207-015-0838-1

Usoskin, I.G., Kovaltsov, G.A., Chatzistergos, T.: 2016b, Dependence of the Sunspot-Group Size on the Level of Solar Activity and its Influence on the Calibration of Solar Observers. *Solar Phys.* **291**, 3793-3805. DOI 10.1007/s11207-016-0993-z

Usoskin, I., Kovaltsov, G., Kiviaho, W.: 2021a, Robustness of Solar-Cycle Empirical Rules Across Different Series Including an Updated Active-Day Fraction (ADF) Sunspot Group Series. *Solar Phys.* **296**, 13, DOI 10.1007/s11207-020-01750-9

Usoskin, I. G., Solanki, S.K., Krivova, N.A., Hofer, B., et al. 2021b, Solar Cyclic Activity over the Last Millennium Reconstructed from Annual 14C Data. *Astron. Astrophys.* **649**, A141. https://doi.org/10.1051/0004-6361/202140711

Vaquero, J.M., Trigo, R.M.: 2015, Redefining the limit dates for the Maunder Minimum. *New Astron.* **34**, 120. DOI 10.1016/j.newast.2014.06.002

Vaquero, J.M., Gallego, M.C., Usoskin, I.G., Kovaltsov, G.A.: 2011, Revisited Sunspot Data: A New Scenario for the Onset of the Maunder Minimum. *Astrophys. J.* **731**, L24. DOI 10.1088/2041-8205/731/2/L24

Vaquero, J.M., Kovaltsov, G.A., Usoskin, I.G., Carrasco, V.M.S., Gallego, M.C.: 2015, Level and length of cyclic solar activity during the Maunder minimum as deduced from the active-day statistics. *Astron. Astrophys.* **577**, A71. DOI 10.1051/0004-6361/201525962

Vaquero, J.M., Svalgaard, L., Carrasco, V.M.S., et al.: 2016, A Revised Collection of Sunspot Group Numbers. *Solar Phys.* **291**, 3061. DOI 10.1007/s11207-016-0982-2

Vasyliunas, V.M., Kan, J.R., Siscoe, G.L., Akasofu, S.-I.: 1982, Scaling relations governing magnetospheric energy transfer. *Planet. Space Sci.* **30**, 359. DOI 10.1016/0032-0633(82)90041-1

Velasco Herrera, V. M., Soon, W., Hoyt, D., and Muraközy, J., Group Sunspot Numbers: 2022, A New Reconstruction of Sunspot Activity Variations from Historical Sunspot Records Using Algorithms from Machine Learning. *Solar Phys.* **297**, 1. DOI 10.1007/s11207-021-01926-x

Vokhmyanin, M., Arlt, R., Zolotova, N.: 2020, Sunspot Positions and Areas from Observations by Thomas Harriot. *Solar Phys.* **295**, 39. DOI 10.1007/s11207-020-01604-4

Vokhmyanin, M., Arlt, R., Zolotova, N.: 2021, Sunspot Positions and Areas from Observations by Cigoli, Galilei, Cologna, Scheiner, and Colonna in 1612 – 1614. *Solar Phys.* **296**, 4. DOI 10.1007/s11207-020-01752-7

Waldmeier, M.: 1938, Chromosphärische Eruptionen. I. *Z. Astrophys.* **16**, 276.